\documentstyle[12pt,epsfig]{report}
\def\thebibliography#1{\leftline{\Large\bf References}\list
  {[\arabic{enumi}]}{\settowidth\labelwidth{[#1]}\leftmargin\labelwidth
    \advance\leftmargin\labelsep
    \usecounter{enumi}}
    \def\newblock{\hskip .11em plus .33em minus .07em}
    \sloppy\clubpenalty4000\widowpenalty4000}

\newcommand{\be}{\begin{eqnarray}}
\newcommand{\ba}{\begin{array}}
\newcommand{\ea}{\end{array}}
\newcommand{\ee}{\end{eqnarray}}
\newcommand{\fp}{F^\prime}
\newcommand{\fpt}{F^{\prime2}}
\newcommand{\sF}{{\rm sin}F}
\newcommand{\sFt}{{{\rm sin}^2}F}
\newcommand{\cF}{{\rm cos}F}
\newcommand{\cFt}{{{\rm cos}^2}F}

\newcommand{\ft}{f_\pi^2}

\newcommand{\gp}{G^\prime}
\newcommand{\gpt}{G^{\prime2}}
\newcommand{\go}{\gamma_1}
\newcommand{\gw}{\gamma_2}
\newcommand{\gt}{\gamma_3}
\newcommand{\wo}{\omega}
\newcommand{\wop}{\omega^\prime}
\newcommand{\wpt}{\omega^{\prime2}}
\newcommand{\alp}{\alpha^\prime}

\newcommand{\bep}{\beta^\prime}

\newcommand{\smalllineskip}{\baselineskip=12pt}

\newcommand{\Ts}{\textstyle}
\newcommand{\Ss}{\scriptstyle}

\newcommand{\vecbf}[1]{\mbox{\boldmath$#1$\unboldmath}}
\newcommand{\vecbfs}[1]{\mbox{\boldmath$\scriptstyle#1$\unboldmath}}

\newcommand{\dslash}{\partial \hskip -0.5em /}
\newcommand{\Aslash}{A \hskip -0.5em /}

\newcommand{\ID}{\mbox{{\sf 1}\hspace{-0.55mm}\rule{0.04em}{1.53ex}}}
\newcommand{\Zn}{\mbox{{\sf Z}\hspace{-3.0mm}
{\raisebox{0.8mm}{$\scriptstyle{\sf/}$}}}\hspace{1.0mm}}
\newcommand{\exeq}{\stackrel{\scriptscriptstyle\raisebox{0.7mm}{!}}
{\scriptscriptstyle \raisebox{0.3mm}{$=$}}}

\begin{document}
\rightline{SU--4240--707}
\rightline{MIT--CTP--2880}
~
\vskip1cm
\Huge
\begin{center}
{\bf The Skyrme Model for Baryons}$^{{\scriptstyle\#}{\textstyle*}}$
\end{center}
\normalsize
\vskip0.5cm
\begin{center}
J. Schechter $^{\textstyle\rm a}$~~~and~~~
H. Weigel$^{\textstyle\dagger,{\rm b}}$\\~\\
$^{\textstyle\rm a)}$Department of Physics, Syracuse University\\
Syracuse, NY 13244--1130\\~\\
$^{\textstyle\rm b)}$Center for Theoretical Physics\\
Laboratory of Nuclear Science and Department of Physics\\
Massachusetts Institute of Technology\\
Cambridge, Ma 02139
\end{center}
\vskip0.5cm
\centerline{\bf ABSTRACT}
\vskip 0.3cm
\centerline{\parbox[t]{15.0cm}
{\baselineskip22pt
We review the Skyrme model approach which treats baryons as solitons of an
effective meson theory. We start out with a historical introduction
and a concise discussion of the original two flavor Skyrme model and
its interpretation. Then we develop the theme, motivated by the large
$N_C$ approximation of QCD, that the {\it effective} Lagrangian of QCD is
in fact one which contains just mesons of all spins. When this Lagrangian
is (at least approximately) determined from the meson sector it
should then yield a zero parameter description of the baryons. We next
discuss the concept of chiral symmetry and the technology involved
in handling the three flavor extension of the model at the collective
level. This material is used to discuss properties of the light baryons
based on three flavor meson Lagrangians containing just pseudoscalars
and also pseudoscalars plus vectors. The improvements obtained by
including vectors are exemplified in the treatment of the {\it proton 
spin puzzle}.}}

\vfill
\noindent
--------------

\noindent
$^{\scriptstyle\#}$\hskip-0.03cm
{\footnotesize Invited review article for INSA--Book--2000.}

\noindent
$^{\textstyle*}$\hskip-0.03cm
{\footnotesize This work is supported in parts by funds provided by the
U.S. Department of Energy (D.O.E.) under cooperative research
agreements \#DR--FG--02--92ER420231 \& \#DF--FC02--94ER40818 and the Deutsche
Forschungsgemeinschaft (DFG) under contracts We 1254/3-1 \& 1254/4-1.}\\
$^{\textstyle\dagger}$\hskip-0.02cm
{\footnotesize {Heisenberg--Fellow}}
\eject

\stepcounter{chapter}
\leftline{\Large \bf 1. Historical background and motivation}
\medskip

The Skyrme model was born around 1960 in a series of increasingly
more detailed papers \cite{sky1}. At that time the prevailing dynamical
model of nuclear forces was that of Yukawa which had been formulated in
the 1930's. Still to come was the concept of fractionally charged quarks
and much further in the future was the recognition that the correct theory
of strong interactions binds these quarks together with non--Abelian
(color)  gauge fields.

In the Yukawa theory, of course, the nucleons are introduced as
fundamental fermion fields while the spin zero pion fields are postulated
to provide the ``glue'' which binds protons and neutrons into nuclei. This
model is acknowledged to work reasonably well as a description of the
long range interactions of nucleons and the prediction of the existence
of pions has been amply confirmed.  

Skyrme's innovation was to provide a model in which the fundamental
fields consisted of just the pions. The nucleon was then obtained,
in the initial approximation, as a certain classical configuration
of the pion fields. The seeming contradiction of making fermi fields
out of bose fields was avoided by arranging the classical field
configuration to possess a non--zero ``winding number''. In modern language
this ``Skyrmion'' is an example of a topological soliton. Such
objects are solutions to the classical field equations with
localized energy density \cite{raj}. They play an important role 
nowadays in many areas of physics and the papers of Skyrme are 
justifiably recognized as pioneering milestones in this
development.

The years following this original idea saw the particle physics
community actively investigating the approaches of quark models, flavor
symmetry, current algebra, chiral dynamics, dual resonance models
and finally color gauge theory to the problem of strong dynamics.
Evidently Skyrme's model was lost in the rush. However the novelty of
the model did stimulate a few interesting papers \cite{will, pt,
bnrs,anw} before the more recent wave of activity in the area.

At first glance, it might appear that a Lagrangian model built out
of only pion fields could not be more different as a description of the 
nucleons from the current picture of three ``valence'' quarks containing
a trivalent ``color'' index and bound together through their interaction
with $SU(3)$ gauge fields. Remarkably, it has turned out that the Skyrme
model is in fact a plausible approximation to this QCD picture. This 
may be understood as follows.

In QCD the gauge coupling constant has an effective strength which 
decreases for interactions at high energy scales (asymptotic freedom) 
but which increases at the low energy scales which are relevant when 
one considers the binding of quarks into nucleons and other hadrons. 
Thus the application of standard perturbation theory techniques to 
the problem of low energy interactions is not expected to be reliable 
and in fact has not produced definitive results. A natural alternative 
approach which retains the possibility of using perturbation theory 
is to imagine that the strong underlying gauge couplings bind
the quarks into particles which may possibly interact with each
other relatively weakly. At low energies these particles should evidently
comprise the pseudoscalar meson fields (pions when restricted to two
``flavors''). Then it is necessary to formulate some {\it effective} 
Lagrangian for the pions. Certainly the Lagrangian should be restricted 
by the correct symmetries of the underlying gauge theory. These must 
include an (approximate) $SU(N_f)$ flavor symmetry~\cite{SU(3) flavor},
where $N_f$ is the number of light flavors.

But there is another symmetry which plays a crucial role. At about the 
same time that Skyrme was contemplating the model under discussion the 
correct formulation \cite{SM et al} of the structure of the effective 
weak (beta--decay etc.) interaction was discovered. It was noted that 
this interaction treated the left and right handed components of fermions
on a completely separate basis. If this distinction is maintained at the
level of the strong interactions one should impose a ``chiral'' left
handed $SU(3)$ flavor $\times$ right handed $SU(3)$ flavor symmetry on the 
effective low energy Lagrangian of mesons. A consequence of this larger 
symmetry is that the meson multiplets must contain scalar as well as the 
desired low--lying pseudoscalar particles. This seemed a bit of an 
embarrassment in that the pseudoscalars are very light while the scalars 
were not well established and presumably heavy. Hence the possibility of a
degenerate symmetry multiplet is implausible. Nevertheless it was realized
\cite{NJL} that the situation was likely to be similar to that met in the 
BCS theory of superconductivity in which the vacuum (ground state) is 
energetically favored to exist in a non--symmetric state. This
``spontaneous breakdown'' picture predicts, in the absence of a needed
small explicit symmetry breaker, exactly zero mass for the pseudoscalars
at the same time that the scalars are massive. In fact it may be
formulated using a so called ``non--linear realization of chiral symmetry''
in such a way that the scalars do not appear at all \cite{gl et al}.
The prototype Lagrangian density for this picture is
\be
{\cal L}=\frac{f_\pi^2}{4}\, {\rm tr}\,
\left(\partial_{\mu}U\partial^{\mu}U^{\dagger}\right)+...\, ,
\label{nlsigma}
\ee
where $U={\rm exp}(\sqrt{2}i\phi/f_\pi)$, $\phi$ being the $3\times3$ 
matrix of the ordinary pseudoscalar mesons and $f_{\pi}=93{\rm MeV}$ 
the ``pion decay constant''. $U$ is a unitary matrix which transforms 
``linearly'' under the chiral transformations. Possible higher derivative 
and symmetry breaking terms have not been explicitly written here. It was 
demonstrated a long time ago that just this term compactly summarizes the 
low energy scattering of pseudoscalar mesons \cite{cro}. Improvements to 
this term form the basis of the ``chiral perturbation theory approach'' 
\cite{chpt}. Now it is believed that a picture like this is expected from 
fundamental QCD. However the same Lagrangian was earlier written by Skyrme 
(in the two rather than three flavor case) in order to explain the
nucleons \cite{sky1} before the present justifications for it were known.

Even in the framework of the chiral Lagrangian given above it would
seem that there is no special {\it a priori} reason not to explicitly add
baryons in a chiral symmetric manner rather than to build them
out of the mesons. Indeed there have been many papers over very many years
which do just this with reasonable phenomenological results \cite{Be95}.
Nevertheless there is an indication from fundamental QCD that the 
soliton treatment of the baryon is more natural. This arises from an 
attempt \cite{t'h} to consider $1/N_C$, the inverse of the number of 
colors in the gauge theory, as a possible expansion parameter for QCD 
which might be meaningful even at low energies. In this approach the 
product $g^{\prime2}=g^2N_C$, where $g$ is the gauge theory coupling 
constant, is held constant. 't Hooft \cite{t'h} showed that for large 
$N_C$ QCD may be considered as a theory of mesons weakly interacting in 
the sense that scattering amplitudes or quadrilinear coupling constants 
are of order $g_{\rm eff}=1/N_C$. Since the baryon mass must start out 
proportional to $N_C$ (noting that the baryon in the $N_C$ model is 
made of $N_C$ quarks) it means that the predicted expressions for baryon 
masses should start out as the inverse of this coupling constant. In 
the framework of the (non--relativistic) mean field treatment 
Witten \cite{Wi79} not only pointed out that the baryon masses indeed 
grow linearly with $N_C$ but also that baryon radii and meson--baryon 
scattering amplitudes are of the order $N_C^0$ while baryon--baryon 
scattering is of ${\cal O}(N_C)$. He in particular recognized that this 
inverse behavior with $g_{\rm eff}$ is just the usual signal that the 
baryon state in question is a soliton of the effective meson theory. 

Naturally, one wonders how these ``modern'' justifications for the
Skyrme approach relate to Skyrme's original motivations. We are fortunate
in having available a reconstructed talk on just this topic by
Skyrme \cite{sky2}. He mentioned three motivations: 1) The idea of unifying
bosons and fermions in a common framework. 2) The feeling that point
particles are inconsistent in the sense that their quantum field theory 
formulation introduces infinities which are only ``swept under the rug'' 
by the renormalization process. 3) The desire to eliminate fermions from a
fundamental formulation since fermions have no simple classical analog.
What seems more fascinating is his awareness that there were probably
some ``hidden'' influences pushing him toward the soliton picture. Directly,
these came from his fascination with Kelvin's idea that the various
atoms should correspond to vortices of different connectivities in some
underlying liquid. In turn, his interest in Kelvin was sparked at an early
age by the presence of a tide prediction machine, designed by Kelvin
and constructed by his great--grandfather, still occupying space in his
great--grandfather's house. An interesting account of this aspect is given
in a paper of Dalitz \cite{Dalitz}.    

Thus it seems that Skyrme's motivations were not those currently used
to justify his model. In particular it appears that he did not choose 
his Lagrangian model to describe spontaneous breakdown of chiral 
symmetry. Rather the non--linear form was chosen to insure that the 
pions were ``angular'' variables which would give multi--valued functions; 
the crossing of different sheets of these functions might then correspond 
to singularities which would realize the baryons. The evident ``moral'' of 
this historical discussion is just that interesting ideas have an uncanny 
way of turning out to be useful and true. In this spirit, we would like 
to continue with the application of Skyrme's ideas to current research
on baryon physics, making use of current motivations but trying
to avoid getting enslaved by them.

\bigskip
\stepcounter{chapter}
\leftline{\Large \bf 2. The Skyrme model for two flavors}
\medskip

In this section we will present the basic technology of the Skyrme model
for baryons. The starting point for the construction of a soliton solution
is the non--linear sigma model Lagrangian (\ref{nlsigma}) already 
introduced in the previous section. As we require a finite energy 
density the chiral field $U$ must approach a constant value at
spatial infinity. We are free to choose this to be unity, {\it i.e.},
\be
U(\vecbf{r},t)\; 
\raisebox{-0.2cm}{${\stackrel{\displaystyle\longrightarrow}
{\scriptscriptstyle |\raisebox{-0.4mm}{$\vecbfs{r}$}
|\to\infty}}$}\; 1 \; .
\label{rlimit}
\ee
This can be considered a mapping from compactified 
coordinate space, a three--sphere $S^3$, to the space which 
is described by the unitary, unimodular matrix $U$, namely  $SU(N_f)$,
where $N_f$ denotes the number of flavors. In the case of two 
flavors the target space is isomorphic to $S^3$. 
The mappings $S^3\to S^3$ fall into distinct equivalence classes.
This signals the existence of soliton configurations because
members of different classes cannot be continuously transformed
into one--another. The equivalence classes are characterized by the 
winding number. This number counts the coverings of the target space
and is the charge $\int d^3x B_0$  associated with the topological 
current
\be
B_\mu=\frac{1}{24\pi^2}\,\epsilon_{\mu\nu\rho\sigma}\,
{\rm tr}\left[\left(U^\dagger\partial^\nu U\right)
\left(U^\dagger\partial^\rho U\right)
\left(U^\dagger\partial^\sigma U\right)\right]\, .
\label{topcur}
\ee
When later discussing the three flavor case we will see that 
this topological current indeed equals the baryon number current.

Although these topological considerations allow the
existence of soliton solutions it turns out that the dynamics
of (\ref{nlsigma}) do not lead to static stable classical solutions. This
can be deduced from a simple consideration, known as Derrick's 
theorem~\cite{De64}. Assume $U_0(\vecbf{r})$ to be such a 
solution. The static energy of $U_0(\lambda\vecbf{r})$, obtained from the
Hamiltonian of (1.1), 
would then be 
\be
E_{\rm cl}^{({\rm nl}\sigma)}[U_0(\lambda\vecbf{r})] = \frac{1}{\lambda}
E_{\rm cl}^{({\rm nl}\sigma)}[U_0(\vecbf{r})]
\label{scal1}
\ee
which does not have a minimum at $\lambda=1$, in contradiction to 
the assumption.  In order to obtain stable solitons Skyrme added a term
to the Lagrangian which is of fourth order in the derivatives, 
\be
{\cal L}^{(\rm Sk)}=\frac{1}{32e^2}\, {\rm tr}\left(
\left[U^\dagger\partial_\mu U,U^\dagger\partial_\nu U\right]
\left[U^\dagger\partial^\mu U,U^\dagger\partial^\nu U\right]
\right)\, .
\label{Skterm}
\ee
Here $e$ is the dimensionless ``Skyrme constant".
Although this term is quartic in the derivatives it was arranged to be at
most
quadratic in the time--derivatives. This makes the
quantization feasible. It is now apparent that a scaled 
configuration may well lead to a minimum of the energy
functional 
\be
E_{\rm cl}^{({\rm tot})}[U_0(\lambda\vecbf{r})] = 
\frac{1}{\lambda}E_{\rm cl}^{({\rm nl}\sigma)}[U_0(\vecbf{r})]
+\lambda E_{\rm cl}^{({\rm Sk})}[U_0(\vecbf{r})]
\label{scal2}
\ee
at $\lambda=1$ provided the configuration $U_0(\vecbf{r})$ 
satisfies $E_{\rm cl}^{({\rm nl}\sigma)}[U_0(\vecbf{r})]=
E_{\rm cl}^{({\rm Sk})}[U_0(\vecbf{r})]$.

{\it A priori} the Euler--Lagrange equations of motion for the 
chiral field $U_0(\vecbf{r})$ are highly non--linear partial differential
equations. To simplify these equations Skyrme adopted the 
famous hedgehog {\it ansatz}
\be
U_0(\vecbf{r})={\rm exp}\left(
i\vecbf{\tau}\cdot\hat{\vecbf{r}} F(r)\right),
\label{hedgehog}
\ee
where $\vecbf{\tau}$ represents the Pauli matrices. This form may 
actually be traced back to the old ``strong coupling" theory \cite{Pa46}. 
Upon substitution of this {\it ansatz} the energy functional
turns into a simple integral involving only the radial 
function $F(r)$,
\be
E[F]=\frac{2\pi f_\pi}{e}\int_0^\infty dx\, \Bigg\{
\left(x^2\fpt+2\sFt\right)+
\sFt\left(2\fpt+\frac{\sFt}{x^2}\right)\Bigg\}\, .
\label{Skeng}
\ee
~\vskip-0.2cm
\noindent
\parbox[l]{9.2cm}{Henceforth this radial function will be called the 
chiral angle. In eq (\ref{Skeng}) a prime indicates a derivative with 
respect to the dimensionless coordinate $x=ef_\pi r$. In this manner
we have completely extracted the dependence on the model parameters.
Imposing $F(\infty)$=0 and noting that $\int d^3r B_0 = (F(0)-F(\infty))/\pi$ 
leads to the boundary condition $F(0)=\pi$ for a unit baryon number 
configuration. The profile function depicted in Fig. 2.1
minimizes (\ref{Skeng}) and is obtained numerically. 
The energy obtained by substituting this solution into (\ref{Skeng}) is 
found to be $E=23.2\pi f_\pi/e$~\cite{anw}.}
\parbox[r]{7.6cm}{
\vskip-0.6cm
\centerline{\hskip -0.0cm
\epsfig{figure=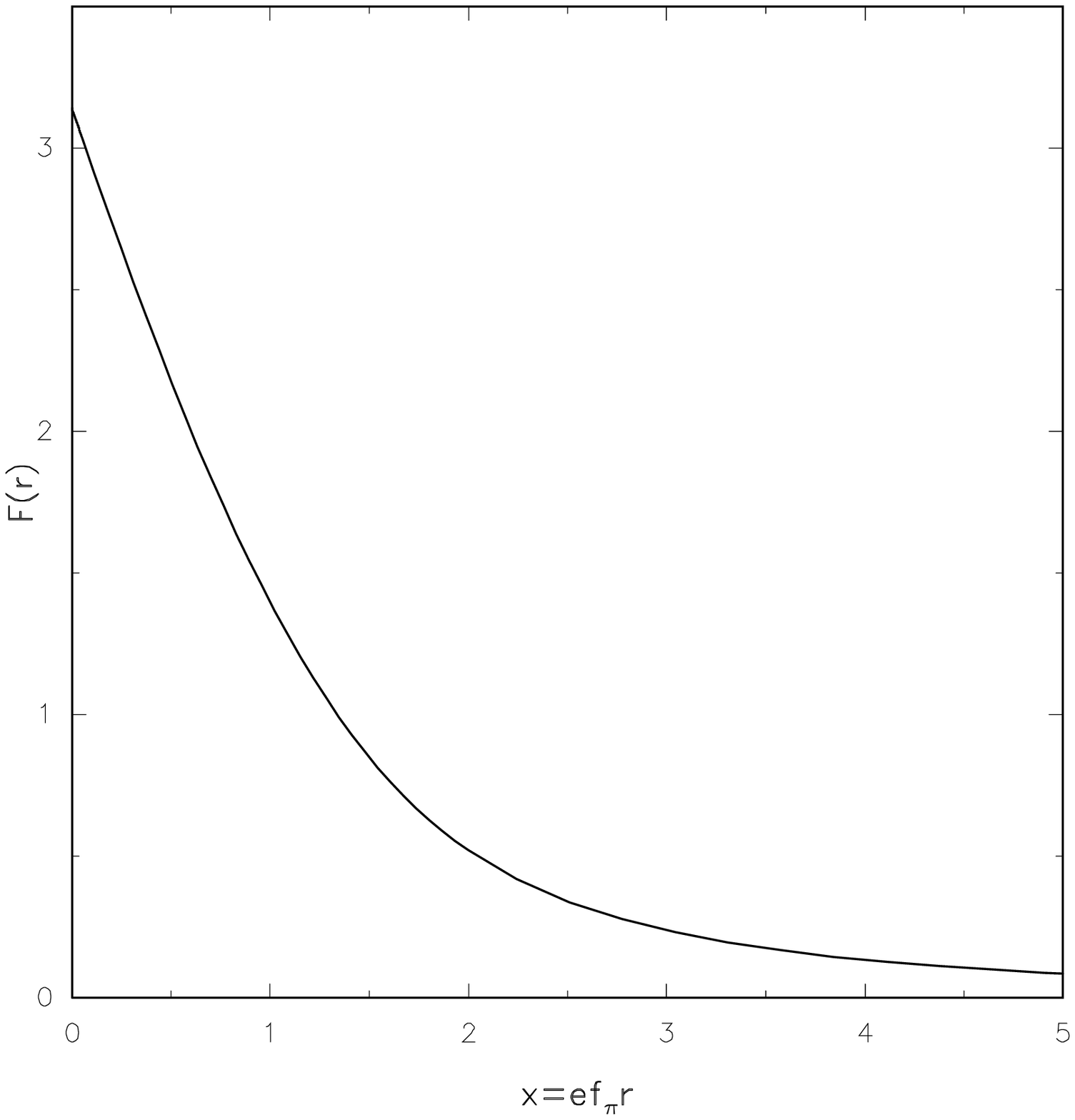,height=5.8cm,width=7.6cm}}
\vskip-0.2cm
\centerline{\hskip -0.0cm
Figure 2.1: Chiral angle}
}

As the {\it ansatz} (\ref{hedgehog}) is not invariant under separate 
spatial or flavor rotations this field configuration does not yet describe 
states with good spin and flavor quantum numbers. As the first step
towards generating such states, time--dependent collective coordinates 
$A(t)$ are introduced which describe the spin and flavor orientation 
of the hedgehog, 
\be
U(\vecbf{r},t)=A(t)U_0(\vecbf{r}\,)A^\dagger(t), \qquad
A(t)\in SU(N_f)\, .
\label{colcor1}
\ee
Note that the hedgehog structure
causes rotations in coordinate and flavor space to be equivalent.
For generality we assumed an arbitrary number of flavors. In the two
flavor case this configuration yields the Lagrange function
\be
L(A,\dot A) = \frac{1}{2}{\alpha^2} \vecbf{\Omega}^2 - E_{\rm cl} 
\label{colcor2}
\ee
where the quantity $\vecbf{\Omega}$ measures the time dependence of the 
collective coordinates
\be
A^\dagger(t) \frac{d}{dt} A(t) = \frac{i}{2}\, 
\vecbf{\tau}\cdot\vecbf{\Omega}\, .
\label{colcor3}
\ee
The constant of proportionality, $\alpha^2$ is computed as 
a spatial integral over the chiral angle to be
$\alpha^2=53.3/(e^3f_\pi)$. Computing the spin as the Noether charge 
of spatial rotations yields $\vecbf{J}={\alpha^2} \vecbf{\Omega}$.
Apparently the system displays all the features of a rigid top. In that 
language $\vecbf{\Omega}$ and $\alpha^2$ are denoted as the angular 
velocity and the moment of inertia, respectively.  

In the second step the collective coordinates are elevated to quantum 
variables. Again this is completely analogous to the quantization of
the rigid top and gives the quantization rule $[J_i,J_j]=i\epsilon_{ijk}J_k$. 
For the hedgehog {\it ansatz} in $SU(2)$ spin and isospin are related via 
the adjoint representation of the collective coordinates, {\it i.e.}
$I_i=-D_{ij}J_j$ with $D_{ij}=(1/2){\rm tr}(\tau_i A \tau_j A^\dagger)$
due to the equivalence of the respective rotations.
Hence only states which have identical spin and isospin are 
allowed in the spectrum. These are the nucleon ($I=J=1/2$) and the
$\Delta$--resonance ($I=J=3/2$). Finally the Hamiltonian for 
the collective coordinates is given by
\be
H_{\rm coll}=E_{\rm cl}+\frac{1}{2\alpha^2}\vecbf{J}\,^2
=E_{\rm cl}+\frac{1}{2\alpha^2}\vecbf{I}\,^2
\label{colcor4}
\ee
which yields the $\Delta$--nucleon mass difference
\be
M_\Delta-M_{\rm N}=\frac{3}{2\alpha^2}\, .
\label{colcor5}
\ee
Using the physical value for the pion decay constant,
$f_\pi=93 {\rm MeV}$ requires us to choose $e\approx4.75$ to 
reproduce the empirical mass difference of $293 {\rm MeV}$.
Substituting $e\approx4.75$ into eq (\ref{Skeng}) yields the 
classical nucleon energy 
$E=23.2\pi f_\pi/e \approx1430 {\rm MeV}$. This is not in
especially good agreement with the experimental value of about 
$939 {\rm MeV}$. However the following points must be kept in mind:

\begin{itemize}
\item[(i)]
The meson Lagrangian consisting of eq (\ref{nlsigma}) plus eq
(\ref{Skterm}) contains only pseudoscalars. We would expect that other
low mass mesons (notably the vectors) should also be included. The large
$N_C$ expansion \cite{t'h, Wi79} requires an infinite number but common
sense suggests a reasonable approximation for explaining hadronic physics
up to about $1{\rm GeV}$ would be to keep those mesons with masses up to 
this value. Certainly the predictions in the mesonic sector of the theory
are noticeably improved by the inclusion of vector mesons. The consistency
of the overall picture requires accurate predictions \underline{both} in 
the mesonic and baryonic sectors of the effective theory.

\item[(ii)]
In nature there are three rather than two ``light" flavors and
this aspect should be included in a realistic formulation. (This
feature also makes more transparent the origin of the topological current
eq (\ref{topcur}).) Furthermore the effects of flavor and chiral symmetry
breaking mediated by the finite values of the quark masses have not yet
been taken into account.

\item[(iii)]
Order of $N_C^0$ corrections to the nucleon mass which have the
structure of the Casimir effect in field theory have also not been
included. These quantum contributions to the energy have been 
estimated to be negative and of the order of a few hundred MeV, 
predicting a total nucleon mass at the order of the experimental 
value \cite{Mo91}. Nevertheless one should be cautious about these 
quantum corrections, after all the Skyrme model is not renormalizable,
leaving a logarithmic scale dependence of the ``renormalized'' Casimir
energy. It seems that at best the quantum corrections can be computed 
in a scenario compatible with the chiral expansion.

\end{itemize}

We will postpone the discussion of a variety of nucleon (and other 
baryons') properties until after we have treated the more general case 
of flavor $SU(3)$. 

To end this section on the basics of the Skyrme model we 
would like to briefly discuss the 
consistency of the Skyrme model with the large $N_C$ picture
of QCD. In section 1 we have already noted that the quadrilinear
coupling between mesons scales like $1/N_C$. To check this
behavior it is convenient to expand the non--linear $\sigma$ model
Lagrangian (\ref{nlsigma}) in powers of the pion field:
\be
\frac{1}{2}\partial_\mu \vecbf{\pi}\cdot \partial^\mu\vecbf{\pi}
+\frac{1}{6f_\pi^2}\left\{
\left(\vecbf{\pi}\cdot\partial_\mu\vecbf{\pi}\right)^2
-\vecbf{\pi}^2\partial_\mu \vecbf{\pi}\cdot 
\partial^\mu\vecbf{\pi}\right\} + 
{\cal O}\left(\vecbf{\pi}^6\right)\, .
\label{expnls}
\ee
Since the quadrilinear coupling constant is $1/f_\pi^2$ we 
deduce that $f_\pi\sim \sqrt{N_C}$. This agrees with general
arguments \cite{Wi79}. Similarly the Skyrme term
(\ref{Skterm}) provides a quartic pion interaction with
the coupling constant $1/(e^2f_\pi^4)$ which implies 
$e\sim 1/\sqrt{N_C}$. Hence the classical energy (\ref{Skeng}) 
grows linearly with the number of colors as asserted 
from the corresponding generalization of QCD. Moreover,
without flavor symmetry breaking large $N_C$ QCD predicts
the baryons of different $J$ to be degenerate \cite{Da94}. This is
perfectly consistent with the mass formula (\ref{colcor4}) because
the moment of inertia also grows linearly with $N_C$ as is indicated 
after eq (\ref{colcor3}), hence the second term in (\ref{colcor4})
behaves like $1/N_C$.

Actually, the understanding of the $N_C$ expansion for baryons
involves some subtleties. Consider the construction of large $N_C$ 
baryons in the quark model. The lowest lying baryons are made of
$N_C$ (taken to be odd) quarks in a totally antisymmetric ({\it i.e.}
singlet) color spin  state with no orbital angular momentum. One expects
particles of all total angular momenta from $J=1/2$ to $J= N_C/2$
to be obtained. In agreement with the spectrum of eq (\ref{colcor4})
we expect $I=J$ for these particles and an infinite number of them as 
$N_C \rightarrow \infty$. The trouble is that there is no experimental 
evidence for any $ I=J=5/2, 7/2 $ etc. particles. This may be
interpreted as evidence that $N_C = 3 $ in nature. Still, the large $ N_C
$ expansion is useful if one computes a quantity which exists in the 
$ N_C = 3 $ theory as a (presumably quickly convergent) Taylor series
in $ 1/N_C $. For the specific case of the higher
excitations $J=5/2,7/2,\ldots$ the above treatment of the rotational
modes seems inadequate because the rotational energy gets as large 
as the classical contribution. By including these modes in the 
Euler--Lagrange equations the widths of these higher excitations have
been estimated to be comparable to their masses \cite{Do94}.  
This makes a particle interpretation of these states problematic
suggesting that they are artifacts of the 
collective quantization method employed rather than of physical relevance.
Possible caveats for these calculations are the instability of these 
configurations against emitting pions \cite{Ba84} and that the results are 
only obtained by analytical continuation in the spin variable.

\bigskip
\stepcounter{chapter}
\leftline{\Large \bf 3. Chiral symmetry and its breaking}
\medskip

In this section we will briefly discuss the concept of chiral 
symmetry which represents a guiding principle for 
extending the Skyrme model. Attention will be limited to those aspects of
this large subject which have direct relevance to the study of Skyrmions.
The basic idea is to construct a model of meson fields which ``mocks up"
as many symmetries and properties of the fundamental QCD Lagrangian as
possible.

\bigskip
\leftline{\large\bf 3.1 The QCD Lagrangian}
\medskip
             Let us first 
recall the matter piece of the QCD Lagrangian
\be
{\cal L}_{\rm QCD}^{\rm matter}&=&
\sum_{f=1}^{N_f} {\bar q}_f \left(i\dslash+g\Aslash-m_f\right)q_f 
\nonumber \\ &=&
\sum_{f=1}^{N_f}\left\{ 
{\bar q}_{f,\rm L}\left(i\dslash+g\Aslash\right)q_{f, \rm L}+
{\bar q}_{f,\rm R}\left(i\dslash+g\Aslash\right)q_{f, \rm R}
-m_f\left({\bar q}_{f,\rm L}q_{f, \rm R}+
{\bar q}_{f,\rm R}q_{f, \rm L}\right)\right\}\, .
\hspace{1cm}
\label{LQCDm}
\ee
Here $A_\mu$ is the matrix representation of the gluon fields
and $g$ the quark--gluon coupling. Most notably we have introduced 
the chiral representation for the QCD current quarks 
\be
q_{\rm L,R} =\frac{1}{2}\left(1\mp\gamma_5\right)q
\label{chiral1}
\ee
of each flavor.

Strictly speaking, the quark mass terms are not part of the QCD Lagrangian
but arise from the Yukawa terms of the full microscopic theory of nature.
A major unsolved problem is to understand the resulting pattern of quark
mass parameters. The phenomenologically determined masses \cite{PDG98}
are $m_u \approx 5{\rm MeV}, m_d \approx 9{\rm MeV}, 
m_s \approx 120 - 170{\rm MeV}, m_c \approx 1.5{\rm GeV}, 
m_b \approx 4.5{\rm GeV}$ and $m_t \approx 175{\rm GeV}$. This
random looking perturbation of the ``strong interaction'' plays a crucial
role in determining the nature of elementary particle physics. In the
region up to about $1{\rm GeV}$ it is not possible to produce particles
containing $c, b$ or $t$ quarks. Then it is usually a good approximation
to simply drop them from the theory. Other approximations are useful 
when dealing with the subspace carrying the flavor quantum number of a 
single ``heavy" quark~\cite{Neubert}. Furthermore in the sector of the 
three ``light" quarks $u,d,s$ it turns out to be fundamental to neglect 
the $u,d,s$ masses as a first approximation and include their effects as 
a perturbation~\cite{GGOR}. This is reasonable because the light masses 
are less than the quantity $\Lambda_{QCD} \approx 250 {\rm MeV}$,
the scale below which the QCD effective coupling gets extremely large.

In the case $m_f=0$ the Lagrangian (\ref{LQCDm}) specialized to the
three light quarks has the global chiral symmetry
\be
U_{\rm L}(3)\times U_{\rm R}(3):\quad
q_{\rm L}\longrightarrow L q_{\rm L}
\quad\quad {\rm and}\quad\quad
q_{\rm R}\longrightarrow R q_{\rm R}\
\quad\quad {\rm with}\quad\quad
q_{\rm L,R}=\pmatrix{q_u\cr q_d\cr q_s}_{\rm L,R}
\label{chsym}
\ee
where $L$ and $R$ are each $3 \times 3$ unitary matrices. Using Noether's
theorem on the classical Lagrangian then yields the conservation of
the eighteen vector and axial vector currents
\be 
j^\mu_{ij} = \bar{q}_j \gamma^\mu q_i
\quad {\rm and}\quad  
j^\mu_{ij,5} = \bar{q}_j \gamma^\mu \gamma_5 q_i,
\label{currents}
\ee
where the latin indices run over $u,d,s$. These currents play an important
role in the theory of weak interactions.

\begin{figure}[ht]
\centerline{
\epsfig{figure=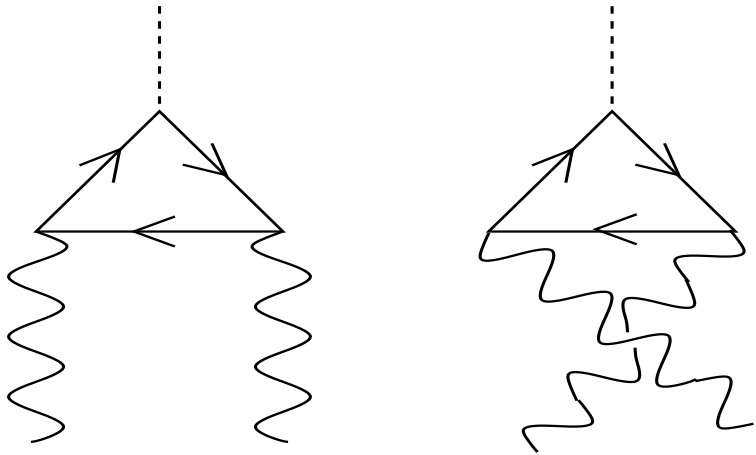,height=6.0cm,width=10.0cm}}
\caption{\label{fig_abj}Adler--Bell--Jackiw anomaly. The arrows indicate
quark lines, the curly lines refer to the gauge bosons and the dashed
lines denote the coupling of the axial singlet current $J_5^\mu$.}
\end{figure}

Now a major discovery of quantum field theory is that consequences of the
classical
field equations of motion  (which can be used to verify the conservation
of the Noether currents) do not necessarily hold at the quantum level.
It is necessary to consider whether there exists a suitable regularization
of the divergent diagrams of the theory which maintains the classical
relations. In the present case, the axial singlet current\footnote{Here
$\lambda^a$, $a=1,\ldots,N_f^2-1$ denote the Gell--Mann 
matrices of $SU(N_f)$ while $\lambda^0$ refers to the
singlet generator.},
$j_{\mu}^5=\bar{q}\gamma_\mu\gamma_5(\lambda^0/2)q$ is not conserved
even for massless quarks. Rather its divergence is proportional to the
gluon field tensor times its dual. This is a result of the well--known
Adler--Bell--Jackiw (ABJ) triangle anomaly \cite{ABJ69} contained in 
the loop--diagrams
shown in figure \ref{fig_abj}. 

The net result is that the true global symmetry of the massless quantum
theory is not $U_{\rm L}(3) \times U_{\rm R}(3)$ but 
$U_{\rm V}(1) \times SU_{\rm L}(3) \times SU_{\rm R}(3)$. The singlet 
vector symmetry, $U_{\rm V}(1)$ corresponds to baryon number conservation. 

A similar situation emerges when external c--number flavor gauge fields
are added to the massless QCD Lagrangian in order to further probe its
structure. Then the so--called non--Abelian anomaly yields non--zero
{\it covariant} derivatives of the $ SU_{\rm L}(3) \times SU_{\rm R}(3)$ 
currents proportional to certain combinations of the corresponding 
external gauge fields \cite{Bardeen}. The non--Abelian anomaly will be 
noted to have important consequences for the theory of Skyrmions.

Although the true symmetry of massless three flavor QCD is 
$SU_{\rm L}(3) \times SU_{\rm R}(3) \times U_{\rm V}(1)$, 
the resulting symmetry of the physical states of the theory is further 
reduced to $ SU_{\rm V}(3) \times U_{\rm V}(1)$ by 
the ``spontaneous breakdown" mechanism. In this mechanism
the vacuum state is not invariant under the full symmetry group. The
massless QCD vacuum is characterized by a non-vanishing ``condensate"
$\langle \bar{q}_u q_u + \bar{q}_d q_d + \bar{q}_s q_s \rangle
\ne 0$.  Under 
an infinitesimal chiral transformation
$L=1+i\sum_{a=0}^{N_f^2-1}\epsilon_{\rm L}^a\lambda^a/2$ and
$R=1+i\sum_{a=0}^{N_f^2-1}\epsilon_{\rm R}^a\lambda^a/2$ the variation
of this quark--bilinear is found to be
\be
\delta \left(\bar{q}q\right)=
i\left(\epsilon_{\rm L}^a-\epsilon_{\rm R}^a\right)
\left(\bar{q}_{\rm L}\frac{\lambda^a}{2}q_{\rm R}-
\bar{q}_{\rm R}\frac{\lambda^a}{2}q_{\rm L}\right)
=\left(\epsilon_{\rm L}^a-\epsilon_{\rm R}^a\right)
\bar{q}\frac{\lambda^a}{2}i\gamma_5 q\, .
\label{delqq}
\ee
Clearly the condensate is invariant only for the subgroup $L=R$ in
eq (\ref{chsym}) which is a vector type transformation. This explains
the physical $ SU_{\rm V}(3) \times U_{\rm V}(1)$ invariance. Note that the
right hand side of eq (\ref{delqq}) represents pseudoscalar objects.
These are ``zero mode" fluctuations of the above vacuum configuration
and the corresponding massless pseudoscalar particles are designated
Nambu--Goldstone bosons. Their scalar chiral ``partners'' -- which would 
be degenerate in mass were it not for the spontaneous symmetry breakdown --
are {\it not} constrained to be massless. This splitting of low--lying
pseudoscalars and scalars expected from massless QCD seems in
qualitative agreement with the experimental situation.

The $ SU_{\rm V}(3) \times U_{\rm V}(1)$ invariance (so--called
``eightfold way'') of massless QCD is, of course, further broken when the
effects of non--zero quark mass terms are included. For later purposes it
is
convenient to rewrite the quark mass terms as:
\be
{\cal L}_{\rm QCD}^{mass}=-\frac{m_u+m_d}{2}\bar{q}{\cal M}q
\qquad {\rm with} \qquad 
{\cal M}=\frac{2+x}{3}\ID+y\lambda_3 +\frac{1-x}{\sqrt 3}\lambda_8.
\label{lsb}
\ee
Here characteristic quark mass ratios are defined by
\be
x=\frac{2m_s}{m_u+m_d}\, ,\quad   y=\frac{m_u-m_d}{m_u+m_d}.
\label{qmr}
\ee
In the limit $y=0$, the theory possesses $SU_{\rm V}(2)$ or
isospin invariance.

\bigskip
\leftline{\large\bf 3.2 Effective Lagrangian of pseudoscalars}
\medskip

The simplest way to mock up low energy QCD is to employ a
$3\times 3$ matrix field $M_{ij}$ which transforms under the chiral group
in the same way as the bilinear quark combination $\bar{q}_{jR}q_{iL}$.
It has the decomposition $M= S + iP$ into hermitian scalar and
pseudoscalar components. A chirally invariant Lagrangian is
\be
{\cal L}= \frac{1}{2}{\rm tr} (\partial_\mu M \partial^\mu M^{\dagger})
-V(M,M^{\dagger}),
\label{lsm}
\ee
where the potential V is a function of invariants like 
${\rm tr}(MM^{\dagger})$, ${\rm tr}(MM^{\dagger}MM^{\dagger})$ etc. 
Spontaneous breakdown to $SU_{\rm V}(3)$ is implemented by choosing V to 
have a minimum such that $\langle M\rangle = const.\times \ID$. Then 
the scalar fields S become massive and can be ``integrated out" by 
imposing a chirally invariant constraint \cite{gl et al}. This represents 
a transition from the linear to the non--linear sigma model. Formally we 
may use the ``polar decomposition" of the matrix $M$ into unitary and 
hermitian factors $M=HU$. Setting $H\rightarrow const.\times \ID$ then 
results in eq (\ref{nlsigma}) again.

In principle, a scenario of this sort can be derived by adding a term
like $\bar{q}_LMq_R + h.c.$ to the QCD Lagrangian and then integrating out
the quark fields\footnote{In the case of QCD this seems to 
be impractical. However, the simpler Nambu--Jona--Lasinio \cite{NJL} 
model for the quark flavor dynamics nicely exemplifies how a
meson functional can be constructed by integrating out quark
degrees of freedom \cite{Eb86}.}. As a result one is left with a
complicated action functional for $M$ and, by further eliminating $H$ as
above, for $U$. Note that $U$ inherits the chiral transformation property of
the quark bilinear: 
\be
U\longrightarrow L U R^\dagger\, .
\label{Utrans}
\ee
It is this identification of the transformation properties which 
provides the important link of the effective chiral theory to QCD 
since it in particular implies that the (Noether) currents 
must be identified. In turn, matrix elements of these currents
will yield the static properties of hadrons.

Of course, in the chiral limit we demand the effective meson 
theory to be strictly invariant under the transformation 
(\ref{Utrans}). Since $U^\dagger U=1$ only derivative terms
can appear and the leading one is just the non--linear $\sigma$
model (\ref{nlsigma}). Clearly also the Skyrme term (\ref{Skterm})
is invariant under (\ref{Utrans}).

At this point one essential ingredient is still missing to ensure
that the chiral field, $U={\rm exp}(i\Phi)$ describes pseudoscalar 
fields, $\Phi$. This requirement demands the parity transformation 
\be
U\,\,
{\stackrel{\raisebox{0.7mm}{$\scriptscriptstyle{\rm parity}$}}
{\textstyle \raisebox{0.3mm}{$\longrightarrow$}}}\,\,
U^\dagger\, .
\label{Uparity}
\ee
However, it is straightforward to verify that the Skyrme model 
Lagrangian is invariant under (\ref{Uparity}) and 
$\vecbf{r}\to-\vecbf{r}$ \underline{separately}. On the level of
the equations of motion we can easily break this unwanted extra 
symmetry by adding a term which contains the Levi--Civita tensor. 
With $\alpha_\mu=(\partial_\mu U)U^\dagger$ we write
\be
\frac{f_\pi^2}{2}\partial_\mu\alpha^\mu + \ldots + 
5\lambda \epsilon_{\mu\nu\rho\sigma}
\alpha^\mu\alpha^\nu\alpha^\rho\alpha^\sigma = 0\, ,
\label{WZ1}
\ee
where the ellipsis refers to the contributions from the Skyrme term 
(\ref{Skterm}) which have the same symmetries as those from the 
non--linear $\sigma$ term (\ref{nlsigma}). Unfortunately, the 
additional term cannot be easily incorporated in the effective Lagrangian
since it does not correspond to the variation of a local term. 
Witten suggested \cite{Wi83} to include it at the level of the
action because the variation of 
\be
\Gamma=\lambda\int_{{\rm M}_5}{\rm tr}\,
\epsilon_{\mu\nu\rho\sigma\tau}
\alpha^\mu\alpha^\nu\alpha^\rho\alpha^\sigma\alpha^\tau
\label{WZ2}
\ee
and the use of Stoke's theorem yields the desired term in the 
equation of motion (\ref{WZ1}) provided the boundary of the five 
dimensional manifold, ${\rm M}_5$ is taken to be Minkowski space, 
{\it i.e.} $\partial {\rm M}_5 = {\rm M}_4$. The choice of 
${\rm M}_5$ is not unique because its complement has the same boundary. 
In order to nevertheless have a unique action the constant 
$\lambda=\frac{-in}{240\pi^2}\, , n\in\Zn$ 
must be quantized\footnote{The reader is referred to the literature
\cite{Wi83} to see the analogy to Dirac's quantization of the 
magnetic monopole.}. It is interesting to study the physical
relevance of (\ref{WZ2}). Expanding in the meson fields $\Phi$ and 
employing again Stoke's theorem reveals that it describes processes
with at least five different pseudoscalars\footnote{For that reason
the term (\ref{WZ2}) vanishes in the case of two flavors which only has 
four different pseudoscalar fields.} like $K^+ K^-\to \pi^+\pi^0\pi^-$. 
As such processes were first discussed by Wess and Zumino \cite{We71} 
who essentially found a power series expression for (\ref{WZ2}), the term 
is commonly named after them\footnote{Some authors refer to the 
Wess--Zumino term in its gauged form {\it i.e.} with external
sources. Unless otherwise noted we will always understand
the Wess--Zumino term to be (\ref{WZ2}).}.

There are further important consequences of the Wess--Zumino 
term (\ref{WZ2}) which can be read off after generalizing it
so its variation with respect to external (electro--weak) gauge 
transformations \cite{Wi83,Ka84} yields the non--Abelian anomaly
\cite{Bardeen}. 
After appropriately
including the corresponding gauge boson fields two striking
features are observed: 
\begin{itemize}
\item[(i)]
A contact interaction for the decay $\pi^0\to \gamma\gamma$ is contained 
in the gauged Wess--Zumino action. On the quark level this process is 
described by the ABJ anomaly involving the diagrams of figure 
\ref{fig_abj} with the external lines representing photons. Identifying
that result with the Wess--Zumino term requires setting $n=N_C$, {\it
i.e.}
the Wess--Zumino term is proportional to the number of colors.
\item[(ii)]
The linear coupling to the $U_{\rm V}(1)$ gauge boson represents the
baryon number current. Indeed it turns out that this current is identical
to the topological current $B_\mu$ in eq (\ref{topcur}).
\end{itemize}

The mocking up of the effects of the $U_{\rm A}(1)$ anomaly in QCD 
involves the $SU(3)$ singlet pseudoscalar particle $\eta^\prime$ and 
will be discussed later when we treat the {\it proton spin puzzle}.

We must also take account of the effects of the finite quark mass 
terms (\ref{lsb}). These transform according to the chiral
$SU_{\rm L}(3) \times SU_{\rm R}(3)$ representation: 
$\vecbf{3}\times\vecbf{3}^*+\vecbf{3}^*\times\vecbf{3}$. 
Note that the matrix $ {\cal M}= {\cal M}^\dagger$ (neglecting the
possibility of strong CP violation) may be considered a ``spurion"
for this transformation property. Then the minimal symmetry breaking piece
 of the effective Lagrangian reads
\be
{\cal L}_{\rm SB}={\rm tr}\,
\left\{{\cal M}\left[-\beta^\prime\left(
\partial_\mu U \partial^\mu U^{\dag}U
+U^{\dag}\partial_\mu U \partial^\mu U^{\dag}\right)
+\delta^\prime\left(U+U^{\dag}-2\right)\right]\right\},
\label{LSB}
\ee
where $\beta^\prime$ and $\delta^\prime$ are two numerical parameters. 
The $\delta^\prime$ term is required to split the pseudoscalar  
meson masses while the $\beta^\prime$ term is required to split the 
pseudoscalar ``decay constants".
The decay constants $f_a$ are defined from the axial vector matrix
elements $\langle 0|j_{5,\mu}^a|\phi_a,p\rangle= i f_a p_\mu$

Working in the isospin invariant limit, the parameters 
$\beta^\prime,\delta^\prime$ and $x$ can then be extracted from the 
knowledge of meson properties \cite{Sch93},
\be
m_\pi^2=\frac{4}{f_\pi^2}\delta^\prime\, , \quad
m_K^2=\frac{4}{f_K^2}\delta^\prime(1+x)\, \quad {\rm and}\quad
\left(\frac{f_K}{f_\pi}\right)^2=
1+\frac{4}{f_\pi^2}\beta^\prime(1-x)\, .
\label{para1}
\ee
This represents the essential input when discussing the Skyrme model 
for three flavors. To sum up, the Lagrangian of only pseudoscalars which
we
shall use for discussing Skyrmions consists of the sum of (\ref{nlsigma}),
(\ref{Skterm}), (\ref{WZ2}) and (\ref{LSB}).

In the chiral perturbation theory approach \cite{chpt} essentially the
most general chirally invariant Lagrangian is written down and ordered
in powers of $\partial\partial\sim{\cal M}$. For example, the leading
terms are eq (\ref{nlsigma}) and the $\delta^\prime$ term of eq (\ref{LSB}).
The next--to--leading terms include:
\be
&&[{\rm tr}(\partial_\mu U\partial^\mu U^{\dag})]^2,\,
{\rm tr}(\partial_\mu U\partial_\nu U^{\dag})
{\rm tr}(\partial^\mu U\partial^\nu U^{\dag}),
\nonumber \\ &&
{\rm tr}(\partial_\mu U\partial^{\mu}U^{\dag}
\partial_\nu U\partial^{\nu}U^{\dag}),\,
{\rm tr}(\partial_\mu U\partial^{\mu}U^{\dag})
{\rm tr}({\cal M}(U+U^{\dag})),\,
{\rm tr}(\partial_\mu U\partial^{\mu}U^{\dag}{\cal M}(U+U^{\dag})),
\nonumber \\ &&
[{\rm tr}({\cal M}(U+U^{\dag})]^2,\,
[{\rm tr}({\cal M}(U-U^{\dag})]^2,
{\rm tr}({\cal M}U^{\dag}{\cal M}U^{\dag}+{\cal M}U{\cal M}U),
\label{8terms}
\ee
each with its own coupling constant. Note that the combinations of terms
which can not be manipulated (by use of various matrix identities)
to become a single trace are suppressed in the $1/N_C$ expansion. This
procedure also entails absorbing the divergent parts of loop corrections
in the coefficients of the listed terms. The result is a joint power
series in energy and the quark masses which can be expected to be very
accurate quite near the $\pi\pi$ threshold in the case of pion--pion
scattering for example. But going higher in energy is very difficult in
this scheme. Furthermore it seems that many of the coefficients mainly
simulate the low energy effects of vector meson exchanges. We shall also
work with a meson Lagrangian which includes the vector particles directly.
This may be thought of as a start on the approach of constructing the
leading (Born) term of the $1/N_C$ expansion, which should include mesons
of all spins.

\bigskip
\leftline{\large\bf 3.3 Effective Lagrangian of pseudoscalars and vectors}
\medskip

We will follow the so--called massive Yang--Mills approach \cite{Ka84}
for introducing the vector meson nonet into the Lagrangian of
pseudoscalars in a chirally invariant manner. In this approach both
vector and axial vector fields are formally introduced as gauge fields
(yielding invariance under {\it local} chiral transformations). Then
globally invariant mass--type terms which break the local chiral invariance
are included. Finally, as in the transition from the linear to the
non--linear sigma model discussed in section 3.1 above, the (heavier)
axial vector mesons are eliminated by a chirally invariant constraint.

We introduce two multiplets with spin one, $A^{\rm L}_\mu$ and 
$A^{\rm R}_\mu$ which we demand to transform under (\ref{chsym}) as 
left-- and right--handed fields, respectively,
\be
A^{\rm L}_\mu\longrightarrow 
L\left(A^{\rm L}_\mu+\frac{i}{g}\partial_\mu\right)L^\dagger
\quad {\rm and} \quad
A^{\rm R}_\mu\longrightarrow 
R\left(A^{\rm R}_\mu+\frac{i}{g}\partial_\mu\right)R^\dagger\, .
\label{Vtrans}
\ee
This allows us to define a covariant derivative for the 
chiral field and field tensors,
\be 
D_\mu U &=&\partial_\mu U -ig A_\mu^{\rm L} U + ig U A_\mu^{\rm R}\, ,
\label{covder} \\
F_{\mu\nu}^{\rm L,R}&=&\partial_\mu A_\nu^{\rm L,R}-
\partial_\nu A_\mu^{\rm L,R}
-ig\left[A_\mu^{\rm L,R},A_\nu^{\rm L,R}\right]\, ,
\label{Ftensor}
\ee
which transform homogeneously under (\ref{chsym}). The chirally 
invariant terms with a minimal number of derivatives read
\be
{\rm tr}\left[(D_\mu U)^\dagger D^\mu U\right]\, ,\quad
{\rm tr}\left[F_{\mu\nu}^{\rm L,R}F^{{\rm L,R},\mu\nu}\right] 
\quad{\rm and}\quad
{\rm tr}\left[F_{\mu\nu}^{\rm L}UF^{{\rm R},\mu\nu}U^\dagger\right]\, .
\label{chinv1}
\ee
In addition we can have mass--type terms for the vector mesons 
\be
{\rm tr}\left[A_\mu^{\rm L}A^{{\rm L},\mu}+
A_\mu^{\rm R}A^{{\rm R},\mu}\right]
\quad{\rm and}\quad
{\rm tr}\left[A_\mu^{\rm L}UA^{{\rm R},\mu}U^\dagger\right]\, ,
\label{chinv2}
\ee
which are still invariant under global chiral transformations.
Of course, many more terms with higher derivatives could be 
written down at the expense of more undetermined parameters.
Now, it is our aim to construct an effective model for 
the vector mesons only; at present we are not interested in the
axial--vector mesons. We have to find a mechanism to 
eliminate the latter without violating the chiral symmetry. 
This can be accomplished by choosing a special ``gauge''
for the vector fields $A_\mu^{\rm L,R}$,
\be
\tilde{A}_\mu^{\rm L}=
\xi\left(\rho_\mu+\frac{i}{g}\partial_\mu\right)\xi^\dagger
\quad{\rm and}\quad
\tilde{A}_\mu^{\rm R}=
\xi^\dagger\left(\rho_\mu+\frac{i}{g}\partial_\mu\right)\xi\, .
\label{defrho}
\ee
Here $\rho_\mu$ is a matrix field with $N_f^2$ components. 
For example, in the case of two flavors it includes both 
the $\rho$ and $\omega$ mesons via
$\rho_\mu=\vecbf{\rho}_\mu\cdot\vecbf{\tau}+\omega_\mu\ID$.
In the case of three flavors, this matrix field is supplemented by
the $K^*$ and $\phi$ mesons. Most importantly we have introduced
the ``square root'', $\xi$ of the chiral field, $U=\xi\xi$ which yields
the chirally invariant relation 
\be
\tilde{A}_\mu^{\rm L}=U\left(\tilde{A}_\mu^{\rm R}
+\frac{i}{g}\partial_\mu\right)U^\dagger\, .
\label{chconst}
\ee
It is actually this so--called unitary constraint which eliminates the 
axial--vector fields in favor of the vector fields $\rho$ without 
spoiling chiral symmetry.

It is interesting to study the behavior of the $\rho$ meson under
chiral transformations. To start off, we recognize that the transformation
of $\xi$ introduces the matrix $K$ which is defined by \cite{Ca69}
\be
\xi\longrightarrow L\xi K^\dagger \exeq K \xi R^\dagger\, .
\label{Kdef}
\ee
Clearly this leaves the transformation law of the chiral 
field (\ref{Utrans}) unchanged. Note that in general the matrix 
$K$ is a position dependent quantity because of $\xi$. 
Demanding now the symmetry transformation
\be
\rho_\mu\longrightarrow K \left(\rho_\mu
+\frac{i}{g}\partial_\mu\right) K^\dagger
\label{Rtrans}
\ee
causes the fields $\tilde{A}^{\rm L,R}$ to transform exactly like
left-- and right--handed vector fields.

Within the unitary gauge the various terms listed above in (\ref{chinv1})
and (\ref{chinv2}) are no longer independent. Introducing the 
homogeneously transforming combinations
\be
p_\mu = \partial_\mu \xi \xi^\dagger + \xi^\dagger \partial_\mu \xi
\quad {\rm and} \quad
R_\mu = \rho_\mu +\frac{i}{2g}\left(\partial_\mu \xi \xi^\dagger 
- \xi^\dagger \partial_\mu \xi\right)
\label{notation}
\ee
the terms up to two derivatives can be combined to a chirally invariant 
Lagrangian of vectors (and pseudoscalars)
\be
{\cal L}_{\rm VM}&=&{\rm tr}\left[
-\frac{1}{4} f_\pi^2p_\mu p^\mu
-\frac{1}{2}F_{\mu\nu}(\rho)F^{\mu\nu}(\rho)
+m_\rho^2R_\mu R^\mu\right]\, ,
\label{LVM}
\ee
where we have used the fact that the coefficient of the term quadratic 
in the $\rho$ meson field is related to the vector meson mass 
$m_\rho=770{\rm MeV}$. Upon expanding the square--root field $\xi$ in powers
of the pseudoscalar
field, one finds that the Lagrangian (\ref{LVM}) contains the 
$\rho\pi\pi$ coupling,
\be
{\cal L}_{\rho\pi\pi}=\frac{m_\rho^2}{2gf_\pi^2}\, \vecbf{\rho}_\mu\cdot
\left(\vecbf{\pi}\times\partial^\mu\vecbf{\pi}\right)\, ,
\label{rpp}
\ee
which can be utilized to fix the coupling constant $g\approx 5.6$ from 
the known decay--width of the process $\rho\to\pi\pi$.

Terms which involve the Levi--Civita tensor $\epsilon_{\mu\nu\rho\sigma}$
are also of great interest for the Skyrme model. For their presentation 
it is most useful 
to introduce the notation of differential forms: $A^R=A_\mu^R dx^\mu,\
d=\partial_\mu dx^\mu$, etc. . Since the left-- and right--handed 
``gauge fields" are related via the unitary constraint (\ref{chconst}) the
number of linearly independent terms, which transform properly under
 chiral transformation as well as parity and charge conjugation,
is quite limited \cite{Ka84}
\be
A^L\alpha^3 \ , \quad
dA^L\alpha A^L-A^L\alpha dA^L+A^L\alpha A^L\alpha \ , \quad
2\left(A^L\right)^3\alpha+\frac{i}{g}A^L\alpha A^L\alpha\ .
\label{anom1}
\ee
For convenience we have again made use of
$\alpha_\mu=(\partial_\mu U)U^\dagger=\xi p \xi^{\dag}$.
Of course, including these terms in the model Lagrangian will
introduce three more parameters: $\gamma_1,\gamma_2$ and $\gamma_3$.
A suitable presentation of this part of the action is given 
in terms of $p$ and $R$ (employing again the notation of differential
forms)
\be
\Gamma_\epsilon&=&\Gamma_{\rm WZ}+\int_{M_4}{\rm tr}\left(
\frac{1}{6}\left[\gamma_1+\frac{3}{2}\gamma_2\right]Rp^3
-\frac{i}{4}g\gamma_2 F(\rho)\left[pR-Rp\right]
-g^2\left[\gamma_2+2\gamma_3\right]R^3p\right)\ ,
\hspace{1cm}
\label{anom2}
\ee
where $\Gamma_{\rm WZ}$ is given in (\ref{WZ2}).
In ref \cite{Ja88} two of the three unknown constants, $\gamma_{1,2,3}$
were determined from purely strong interaction processes like
$\omega\rightarrow 3\pi$. Defining $\tilde h=-2\sqrt2\gamma_1/3$,
$\tilde g_{VV\phi}=g\gamma_2$ and $\kappa=\gamma_3/\gamma_2$
the central values $\tilde h=\pm0.4$ and $\tilde g_{VV\phi}=\pm1.9$ were
found. Within experimental uncertainties (stemming from the errors
in the $\omega - \phi$ mixing angle) these may vary in the range
$\tilde h=-0.15,\ldots,0.7$ and $\tilde g_{VV\phi}=1.3,\ldots, 2.2$
subject to the condition
$\vert\tilde g_{VV\phi}-\tilde h\vert\approx 1.5$. The third parameter,
$\kappa$ could not be fixed in the meson sector. From studies
\cite{Me89} of nucleon properties in the two flavor model it was
argued that $\kappa\approx1$ represents a reasonable choice.

The sum of the space integral of (\ref{LVM}) and (\ref{anom2}) comprise
the chirally invariant
part of the effective action we shall use for discussing the
soliton in the vector meson model. Note that the second piece of
(\ref{anom2}) can be gauged with external fields \cite{Ja88} so as
to make no contribution to the non--Abelian anomaly. The first piece
 $\Gamma_{\rm WZ}$ then correctly supplies the non--Abelian anomaly.
Furthermore the second piece of (\ref{anom2}) stabilizes the soliton
without the need for including the Skyrme term (\ref{Skterm}).

We must still include the effects of symmetry breaking due to finite
quark masses in the vector meson system.
To leading order in the symmetry breaking, an appropriate term which 
behaves properly under chiral transformations, can be constructed by
analogy to 
the last expression in (\ref{chinv2})
\be
-\alpha^\prime{\rm tr}\left[{\cal M}\left(A_\mu^L U A^{R\mu}+
A_\mu^R U^{\dag} A^{L\mu}\right)\right] .
\label{vmsymbr}
\ee
 This 
leading contribution not only distinguishes between 
the $\rho$ and $K^*$ masses but also contributes to the different 
decay constants of the pseudoscalar mesons via the unitary gauge 
(\ref{defrho}). The reader may consult ref \cite{Sch93} for recent 
discussion of higher order symmetry breaking terms.

We would like to end this section on including vector mesons by 
noting that the same Lagrangian is obtained within the so--called
hidden gauge approach \cite{Ba88}, once the same symmetries are required. 
This shows that these two approaches are in fact identical.

\bigskip
\leftline{\Large\bf 3.4 Other aspects}
\medskip

We expect that baryons should appear as solitons of the large $N_C$
effective meson Lagrangian for any number of flavors $N_f$. In the
case of three (or more) light flavors the Wess--Zumino term guarantees,
as discussed in section 3.1, that the baryon number (\ref{topcur})
is obtained in a self--contained manner from the Lagrangian. This can be
used to check that the soliton indeed has the correct baryon number.

Now in the two flavor case, the same kind of soliton solution exists.
However the Wess--Zumino term vanishes identically so we cannot similarly
check its baryon number in a self--contained way. The situation is even
more peculiar for $N_f = 1$. There the Skyrme model represents a
mapping $S^3 \rightarrow S^1$ which does not contain topologically stable
configurations. However, we are not forced to use an effective Lagrangian
of the same form. In this case it is probably more realistic to construct
the Lagrangian by including isoscalars like the spin--0~$\sigma$--field
and the spin--1~$\omega$--field. Such a Lagrangian might have a soliton 
solution (not necessarily topological) but a check of its baryon number 
may also not be available in a self--contained way. These examples seem to
indicate that the form of the relevant effective Lagrangian may have
a non--trivial $N_f$ dependence (at least for small $N_f$).

Another interesting question, related in the sense of understanding
whether physical features of the solitons can be traced to particular
pieces of the effective Lagrangian, concerns the {\it stabilization}
of the soliton. In section 2 we noted that the Skyrme term (\ref{Skterm})
was introduced precisely for this purpose. There is an often mentioned
``derivation" of this term from the piece of the vector meson Lagrangian
(\ref{LVM}) above which goes as follows. In a large mass expansion, 
$m_\rho\to\infty$ the equation of motion for the vector meson field simply
becomes $R_\mu=0$. Substituting this into the remainder of the vector 
meson Lagrangian (\ref{LVM}) 
$$
F_{\mu\nu}(\rho)\longrightarrow
F_{\mu\nu}\left(\frac{-i}{2g}[\partial_\mu \xi \xi^\dagger
- \xi^\dagger \partial_\mu \xi]\right)
$$
yields exactly the Skyrme term (\ref{Skterm}) with the identification $g=e$. 
Although the numbers 5.6 and 4.75 are in reasonable agreement there is 
one caveat to this appealing derivation of the Skyrme term. While the
Skyrme model does yield stable solitons, however, for arbitrary 
large but finite $m_\rho$ the model (\ref{LVM}) 
does not contain stable soliton solutions. Thus one seems to have achieved
stabilization merely by approximating a model in which stabilization does
not exist. Clearly we have not obtained a ``physical origin" for the
stabilization mechanism. As mentioned in the previous section, the second
piece of (\ref{anom2}) stabilizes the soliton in the vector meson
Lagrangian without a need for the Skyrme term. It is also possible
that, as in the case of the s--wave ground state of hydrogen, stability is
achieved at the quantum, rather than at the classical, level. Several
investigations of this possibility have been made \cite{quantstab}
based on just the non--linear sigma model term (\ref{nlsigma}),
although an assumption on the allowed chiral profiles seems to be
required.

\bigskip
\stepcounter{chapter}
\leftline{\Large \bf 4. The Skyrme model with three flavors}
\medskip

It is well established \cite{SU(3) flavor} that the neutron ($n$) and
proton ($p$) belong to a multiplet with six other members (the iso--singlet
$\Lambda$, the iso--doublet $\Xi$ and the iso--triplet $\Sigma$). To 
try to understand $n$ and $p$ alone is to look at only a small piece
of a large picture. Thus we must consider the three flavor generalization
of the treatment in section 2. First (in the present section) we shall 
consider the Lagrangian of pseudoscalars alone, discussed in section 3.2.
The new features arise from the inclusion of flavor $SU(3)$ symmetry
breaking terms (see (\ref{LSB}) together with (\ref{lsb}) and (\ref{qmr}))
as well as the Wess--Zumino term (\ref{WZ2}). Both of these features
involve non--trivial extensions of the formalism and interesting 
``physics''.

The first step towards including the strangeness degrees of freedom is to 
actually take the chiral field to be a $U(1)\otimes SU(3)$ matrix. To 
be precise, the three flavor chiral field is defined as
\be
U(x)={\rm exp}\left(i\frac{\sqrt{2}}{\sqrt{3}f_\pi}
\eta_0\right)\,
{\rm exp}\left(i\Phi\right).
\label{su3chif}
\ee
While the singlet field $\eta_0$ is separated the matrix field 
$\Phi$ now not only contains the pion degrees of freedom but also 
the kaons and the non--singlet component of the $\eta$ fields,
\be
\Phi=\sum_{a=1}^8\frac{\sqrt{2}}{f_\pi}\phi^a\lambda^a=\pmatrix{
\frac{1}{\sqrt2}\pi^0+\frac{1}{\sqrt6}\eta_8 &
\pi^+ & K^+ \cr
\pi^- & -\frac{1}{\sqrt2}\pi^0+
\frac{1}{\sqrt6}\eta_8 & K^0 \cr
K^- & \bar{K}^0 & 
-\frac{2}{\sqrt6}\eta_8 \cr}.
\label{eightfold}
\ee
Here $\lambda^a$ denote the Gell--Mann matrices. Note that in the 
presence of derivative--type symmetry breakers ({\it e.g.} the 
$\beta^\prime$ term in (\ref{LSB})) the normalization of the 
fields gets shifted; the ``physical'' fields are gotten by multiplying
the fields above by some constants as $Z_\pi \pi^+$, $Z_K K^+$, etc.;
similarly the physical decay constants are $Z_\pi f_\pi=93{\rm MeV}$, 
$Z_K f_K\approx113{\rm MeV}$, etc. For the $Z$'s we have 
\be
Z_\pi=\left(1-\frac{8}{f_\pi^2}\beta^\prime\right)^{\frac{1}{2}}
\, ,\quad
Z_K=\left(1-\frac{4}{f_\pi^2}(1+x)\beta^\prime\right)^{\frac{1}{2}}
\quad {\rm etc.}
\label{psnorm}
\ee

We clearly need a suitable generalization of the Skyrme 
{\it ansatz} (\ref{hedgehog}). It turns out that it is correct to
just embed the $SU(2)$ hedgehog in the $SU(3)$ matrix.
Flavor symmetry breaking implies that field configurations 
which have non--zero strangeness possess a classical energy which
(at least in the unit baryon number sector) is larger than that of
a zero strangeness configuration. Thus we choose the embedding:
\be
U_0(\mbox{\boldmath $r$})=\pmatrix{
\vspace{-1.8mm}
{\rm exp}\left(i\vecbf{\tau}\cdot
\hat{\vecbf{r}}F(r)\right)\hspace{-15pt}&
\raisebox{-1.0mm}{\mbox{\huge $|$}}
&\hspace{-16pt}{\scriptstyle\raisebox{0.4mm}{$0$} \atop 
\scriptstyle\raisebox{-0.4mm}{$0$}}\cr
\vspace{-1.8mm}
-\hspace{-5pt}-\hspace{-5pt}-\hspace{-5pt}-\hspace{-5pt}
-\hspace{-5pt}-\hspace{-5pt}-\hspace{-5pt}-\hspace{-5pt}
-\hspace{-5pt}-\hspace{-5pt}-\hspace{-5pt}-\hspace{-5pt}
-\hspace{-15pt}&-&\hspace{-15pt}
-\hspace{-5pt}-\hspace{-5pt}-\hspace{-5pt}-\cr
\quad 0\qquad 0& \raisebox{1.0mm}{\mbox{\large $|$}}
&\hspace{-14pt} 1\cr}\, .
\label{su3hedgehog}
\ee
Hence the classical energy will not be modified and the soliton
profile, $F(r)$ is that in figure~2.1. The effects
of the strange degrees of freedom are hence visible 
when states with baryon quantum numbers are generated via the
collective coordinate approach. 

The collective coordinate matrix $A(t)$ defined in eq (\ref{colcor1}) is 
now taken from $SU(3)$ and in analogy to eq (\ref{colcor3}), now leads to 
eight angular velocities,
\be
A^\dagger(t) \frac{d}{dt} A(t) = \frac{i}{2}\,
\sum_{a=1}^8 \lambda^a \Omega_a\, .
\label{su3omega}
\ee
In addition to the angular velocities $\Omega_a$ the adjoint 
representation
\be
D_{ab}=\frac{1}{2}\, 
{\rm tr}\left(\lambda_a A \lambda_b A^\dagger\right)
\label{adjrep}
\ee
of the collective rotations, $A(t)$ will be important, in particular
in the context of flavor symmetry breaking.

Substituting $U=A(t)U_0(\vecbf{r})A^\dagger(t)$ into the pseudoscalar
Lagrangian of section 3.2 without the symmetry breaker gives rise 
after a spatial integration to the collective Lagrangian
\be
L_{\rm Skyrme}(A,\dot A) +L_{\rm WZ}(A,\dot A)
= \frac{1}{2}{\alpha^2} \sum_{i=1}^3 \Omega_i^2
  +\frac{1}{2}{\beta^2} \sum_{\alpha=4}^7 \Omega_\alpha^2
  -\frac{N_C B}{2\sqrt{3}}\, \Omega_8-E_{\rm cl}\, .
\label{su3lag}
\ee
The $SU(2)$ moment of inertia $\alpha^2$ remains unchanged while the 
moment of inertia $\beta^2$ for rotations into the strange directions is 
a new functional of the pseudoscalar fields.
The fact that the eighth component of the angular velocity vector
does not appear quadratically in (\ref{su3lag}) is a consequence of 
$[U_0,\lambda_8]=0$. The term proportional to $B\Omega_8$, where $B$ 
is the baryon number arises from $\Gamma_{\rm WZ}$. In order to 
obtain it we make use of the separation \cite{Ba85} 
\be
\Gamma_{\rm WZ}[U]=\Gamma_{\rm WZ}[U_0]
-\frac{iN_C}{48\pi^2}\int d^4x\, {\rm tr}
\left\{\left[(U_0^\dagger d U_0)^3+(U_0 d U_0^\dagger)^3\right]
(A^\dagger d A)\right\}\, ,
\label{WZ3}
\ee
where, again, Stoke's theorem has been employed. As $U_0$ is static
we have $\Gamma_{\rm WZ}[U_0]=0$ and the remainder becomes a local
object which is straightforwardly evaluated.

\bigskip
\leftline{\large\bf 4.1 Quantization of the three flavor 
collective Lagrangian}
\medskip

In order to quantize the three flavor Lagrangian (\ref{su3lag})
we require the operators for spin and flavor as Noether charges. 
As a consequence of the hedgehog structure, the infinitesimal change 
under spatial rotations can be written as a derivative with respect 
to $\mbox{\boldmath $\Omega$}$
\be
\left[\mbox{\boldmath $r$}\times\mbox{\boldmath $\partial$},
U(\mbox{\boldmath $r$},t)\right]=
\frac{\partial \dot U(\mbox{\boldmath $r$},t)}
{\partial \mbox{\boldmath $\Omega$}}\ .
\label{infrot}
\ee
By the Noether construction this leads to the spin operator 
$\mbox{\boldmath $J$}
=\partial L(A,\Omega_a)/\partial\mbox{\boldmath $\Omega$}$.
The quantization of the ``$SU(3)$ rigid top'' proceeds by generalizing 
this result to the so--called right generators 
\be
R_a=-{{\partial}\over{\partial\Omega_a}}
\left(L_{\rm Skyrme}+L_{\rm WZ}\right)=\cases{
-\alpha^2\Omega_a=-J_a,&a=1,2,3\cr
-\beta^2\Omega_a,&a=4,..,7\cr
\frac{N_CB}{2\sqrt3},&a=8} \ .
\label{Rgen}
\ee
The quantization prescription then demands the commutation 
relation $[R_a,R_b]=-if_{abc}R_c$ with $f_{abc}$ being the 
antisymmetric structure constants of $SU(3)$. Explicit 
expressions for these generators in terms of an ``Euler--angle" 
parameterization of $A$ are presented in ref \cite{We96}. The 
so--called left generators, which are defined by the rotation
$L_a=D_{ab}R_b$, satisfy the commutation relations
$[L_a,L_b]=if_{abc}L_c$. They provide the isospin,
$I_i=L_i\ (i=1,2,3)$ and hypercharge, $Y=2L_8/\sqrt3$ operators.

The generator $R_8$ is linearly connected to the so--called right 
hypercharge $Y_R=2R_8/\sqrt3=1$ for $B=1$ and $N_C=3$. In analogy 
to the Gell--Mann Nishijima relation a right charge 
\be
Q_R=-J_3+\frac{Y_R}{2}
\label{GMN}
\ee
may be defined. Completing the analogy we note that the eigenvalues 
of $Q_R$ are $0,\pm1/3$, $\pm2/3$, $\pm1$, $\ldots$. Hence for 
$Y_R=1$ the relation (\ref{GMN}) can only be fulfilled when the 
eigenvalue of $J_3$ is half--integer. This yields the important 
conclusion that the $SU(3)$ model describes fermions. {\it A priori} 
this is not expected since the starting point has been 
an effective model of bosons. This discussion can be generalized
to arbitrary $N_C$ showing that the Skyrmion describes fermions 
when $N_C$ is odd and bosons when $N_C$ is even \cite{Wi83}. This, 
of course, is expected from considering baryons as being composed of 
$N_C$ quarks. We conclude that the proper incorporation of the 
anomaly structure of QCD leads to the desired spin--statistics relation. 

\bigskip
\leftline{\large\bf 4.2 Flavor symmetry breaking and baryon 
spectrum}
\nopagebreak
\medskip
\nopagebreak
For a realistic treatment of baryon states in the space
of the collective coordinates we have to supplement the collective 
Lagrangian by the flavor symmetry breaking pieces associated with 
(\ref{LSB}). Substituting the flavor rotating hedgehog yields the
symmetry breaking piece in the collective Lagrangian,
\be
L_{\rm SB}=-\frac{1}{2}\gamma\left(1-D_{88}\right)
\label{LCSB}
\ee
with the coefficient, $\gamma$ being linear in the symmetry breaking
parameter $1-x$, {\it i.e.}
\be
\gamma=\frac{32\pi}{3}(x-1)\int dr \left\{
\delta^\prime r^2(1-\cF)-\beta^\prime\cF
\left(\fpt r^2+2\sFt\right)\right\}\, .
\label{gasky}
\ee
$D_{88}(A)$ is defined in eq (\ref{adjrep}).
Putting pieces (\ref{su3lag}) and (\ref{LCSB}) together, 
the Hamiltonian for the collective coordinates is obtained
as the Legendre transform $H=-\sum_{a=1}^8R_a\Omega_a-L$
\be
H(A,R_a)&=&E_{\rm cl}
+\frac{1}{2}\left[\frac{1}{\alpha^2}-\frac{1}{\beta^2}\right]
\vecbf{J}^2+\frac{1}{2\beta^2}C_2(SU(3))
-\frac{3}{8\beta^2}
+\frac{1}{2}\gamma \left(1-D_{88}\right)
\label{collham}
\ee
for $B=1$ and $N_C=3$. The constraint $R_8=\frac{\sqrt3}{2}$,
which yielded the spin--statistics relation, commutes with $H$
permitting one to substitute this value. The term involving
$\sum_{\alpha=4}^7R_\alpha^2$ has been re--expressed by introducing the
quadratic Casimir operator of $SU(3)$, $C_2(SU(3))=\sum_{a=1}^8R_a^2$.
The standard $SU(3)$ representations are eigenstates of $C_2(SU(3))$
with eigenvalues $\mu$. For example, the octet representation 
$\vecbf{8}$ has $\mu_{\vecbfs{8}}=3$ while 
$\mu_{\vecbfs{10}}=\mu_{\overline{\vecbfs{10}}}=6$ and
$\mu_{\vecbfs{27}}=8$. These representations diagonalize\footnote{The 
hedgehog structure of the classical configuration $U_0$ constrains the
permissible $SU(3)$ irreducible representations to those which have at 
least one state with $I=J$ \cite{Ma84}.} the collective Hamiltonian in 
the absence of symmetry breaking,~$\gamma=0$. 

Now consider the full collective Hamiltonian including the symmetry
breaking. It seems reasonable to assume these $\gamma=0$ eigenstates as 
a basis to diagonalize the full Hamiltonian.
In a perturbative treatment up to third order in $\gamma$ for the 
$\frac{1}{2}^+$ baryons only the representations $\vecbf{8},
\overline{\vecbf{10}}$ and $\vecbf{27}$ contribute \cite{Pa89}. 
For that reason the perturbative treatment is still simple, although 
one must go beyond leading order. In particular this implies that the
nucleon is no longer a pure octet state but rather contains sizable
admixture of the nucleon type states in higher dimensional 
representations,
\be
|N\rangle=|N,{\bf 8}\rangle
+0.0745\gamma\beta^2|N, {\overline {\bf 10}}\rangle
+0.0490\gamma\beta^2|N, {\bf 27}\rangle+\ldots\ ,
\label{nwfexp}
\ee
where the coefficients of the effective symmetry breaker $\gamma\beta^2$
are computed from $SU(3)$ Clebsch--Gordon coefficients \cite{Sw63}.
The nucleon is seen to have a roughly 25\% amplitude to contain the
$\overline{\vecbf{10}}$ state.

Although this perturbative treatment provides a physical picture of
the symmetry breaking effects it actually turns out that the full
Hamiltonian (\ref{collham}) can be \underline{exactly} diagonalized 
numerically. The important ingredient is that within a suitable
``Euler--angle'' representation of the rotations $A$, the symmetry
breaker $1-D_{88}$ depends only on one of these eight angles. In 
each isospin channel the eigenvalue equation 
\be
\left[C_2(SU(3))+\beta^2\gamma\left(1-D_{88}\right)\right]\Psi=
\epsilon_{\rm SB}\Psi
\label{c2problem}
\ee
then reduces to a set of coupled ordinary differential equations which 
can be integrated numerically. Here we do not wish to discuss this
approach in full detail; rather we refer the reader to the original
work by Yabu and Ando \cite{Ya88} and exhaustive applications of this
method involving the present authors \cite{We90,Pa91,We96}. Having
obtained the eigenvalue $\epsilon_{\rm SB}$ the baryon masses are 
straightforwardly computed from
\be
M_B=E+\frac{1}{2}\left(\frac{1}{\alpha^2}-\frac{1}{\beta^2}\right)
J(J+1)-\frac{3}{8\beta^2}+\frac{1}{2\beta^2}\epsilon_{\rm SB}\ .
\label{bmass}
\ee
As already mentioned this diagonalization procedure is equivalent to 
the perturbation expansion. For small enough symmetry breaking 
$\beta^2\gamma$ even first order is sufficient. In that (unjustified)
case the famous Gell--Mann--Okubo mass formulae \cite{Ok62,SU(3) flavor} 
holds exactly: 
\be
2\left(M_N+M_\Xi\right)&=&M_\Sigma+3M_\Lambda
\label{GMOmass1} \\
M_\Omega-M_{\Xi^*}&=&M_{\Xi^*}-M_{\Sigma^*}=
M_{\Sigma^*}-M_\Delta\, .
\label{GMOmass2}
\ee

Additional corrections \cite{We90} arise when we allow for 
non--zero classical $K$--meson fields to get induced by ``rotations''
$\Omega_\alpha$ into the strange directions. These are energetically
favorable since they maximize the strange moment of inertia $\beta^2$.
With a parameterization 
\be
\pmatrix{K^+\cr K^0}= W(r)\, \hat{\vecbf{r}}\cdot\vecbf{\tau}\,
\pmatrix{\Omega_4-i\Omega_5 \cr\Omega_6-i\Omega_7}\, ,
\label{kind}
\ee
the radial function $W(r)$ is determined from applying a variational
principle to $\beta^2$. In principle one must enforce that the {\it ansatz}
(\ref{kind}) has no overlap with any global rotation of the classical
solution (\ref{su3hedgehog}).

We adjust the only free parameter, $e\approx 4$ to the mass differences 
of the low--lying $\frac{1}{2}^+$ and $\frac{3}{2}^+$ baryons. The 
resulting baryon spectrum is shown in table \ref{ta_mdiffps}.
\begin{table}[ht]
\caption{\label{ta_mdiffps}\small
The mass differences, which are obtained by exact diagonalization
of the collective Hamiltonian (\protect\ref{collham}), of the
${\Ss \frac{1}{2}^+}$ and ${\Ss \frac{3}{2}^+}$ baryons in the 
pseudoscalar model for ${\Ss e=4.0}$ are compared to the experimental 
data. The values in parentheses are obtained by enforcing the zero 
overlap condition mentioned after (\ref{kind}) \protect\cite{We96}.
In that case the Skyrme parameter has slightly been readjusted to 
${\Ss e=3.9}$. All data are in MeV.}
~
\newline
\centerline{\small\smalllineskip
\begin{tabular}{c | c c}
Baryons & Model & Expt. \\
\hline
$\Lambda-N$        & 154 (163) & 177 \\
$\Sigma-N$         & 242 (264) & 254 \\
$\Xi-N$            & 366 (388) & 379 \\
$\Delta-N$         & 278 (268) & 293 \\
$\Sigma^*-N$       & 410 (406) & 446 \\
$\Xi^*-N$          & 544 (545) & 591 \\
$\Omega-N$         & 677 (680) & 733 \\
\end{tabular}}
\end{table}
Apparently the three flavor Skyrme model reasonably accounts for
the empirical mass differences. The original studies 
\cite{Gu84,Pr83,Ch85,Ya88} yielded far too low mass splittings 
between baryons of different strangeness for physically motivated 
parameters of the effective Lagrangian\footnote{Many of these authors 
considered $f_\pi$ as a free parameter fitted to the absolute values 
of the baryon masses. Without the $\beta^\prime$ term this yielded 
$f_\pi$ as low as 25MeV \cite{Pr83}.}. A major reason for the improvement 
is the fact that $\gamma$ is significantly enlarged by including the 
effects associated with $f_K\ne f_\pi$~\cite{We90}. It is also apparent 
from table \ref{ta_mdiffps} that enforcing the zero overlap condition
for the induced kaon components can be compensated by a small variation 
of the Skyrme parameter, $e$. This indicates that possible double 
counting effects play only a minor role. It is interesting to remark that
the mass differences for the $\frac{1}{2}^+$ baryons deviate strongly
from the predictions in leading order of the flavor symmetry breaking.
This can easily be observed from the ratios
\be
\left(M_\Lambda-M_N\right):
\left(M_\Sigma-M_\Lambda\right):
\left(M_\Xi-M_\Sigma\right)=1:0.52:0.85
\label{mratio}
\ee
which are in much better agreement with the experimental data
(1:0.43:0.69) than the leading order result (1:1:0.5). Obviously 
the higher order contributions are important. This also indicates 
that the baryon wave--functions contain sizable admixture of higher 
dimensional $SU(3)$ representations, {\it cf.} eq (\ref{nwfexp}). 
Nevertheless the deviation from the Gell--Mann--Okubo relations 
(\ref{GMOmass1}) is only moderate, in particular the equal spacing 
among the $\frac{3}{2}^+$ baryons is well reproduced. Finally we note
that, as discussed in point (iii) of section 2, the absolute mass of the
nucleon is also too high in the three flavor case. Again we must rely
on the $N_C^0$ corrections mentioned.

\bigskip
\leftline{\large\bf 4.3 Electromagnetic properties of
$\frac{\vecbf{1}}{\vecbf{2}}^{\vecbf{+}}$ baryons}
\medskip

The value for the Skyrme parameter $e=4.0$ obtained from this
best fit to the baryon mass differences is next employed to evaluate 
static properties of baryons within this model. In order to do 
so one first constructs the Noether currents associated 
with the symmetry transformation (\ref{Utrans}). A convenient 
method is to extend these global symmetries to local ones by 
introducing external gauge fields ({\it e.g.} the gauge fields 
of the electroweak interactions) into the total action {\it i.e.}
(\ref{nlsigma}), (\ref{Skterm}), (\ref{WZ2}) and (\ref{LSB}). 
The Noether currents are then read off as the expressions which 
couple linearly to these gauge fields. This procedure is especially 
appropriate for the Wess--Zumino term (\ref{WZ2}) because this 
non--local term can only be made gauge invariant by a lengthy 
iterative procedure \cite{Wi83,Ka84}. The final form of the nonet 
($a=0,\ldots,8$) vector ($V_\mu^a$) and axial--vector ($A_\mu^a$) 
currents reads \cite{Pa91} (for $N_C=3$)
\be
V_\mu^a (A_\mu^a) & = & -\frac{i}{2}f_\pi^2\ 
{\rm tr}\left\{\left(\xi Q^a\xi^\dagger \mp 
\xi^\dagger Q^a\xi\right)p_\mu\right\}
-\frac{i}{8e^2}{\rm tr}\left\{
\left(\xi Q^a\xi^\dagger \mp \xi^\dagger Q^a\xi\right)
\left[p_\nu,\left[p_\mu,p_\nu\right]\right] \right\}
\nonumber \\ &&
-\frac{1}{16\pi^2}\epsilon^{\mu\nu\rho\sigma}
{\rm tr}\left\{\left(\xi Q^a\xi^\dagger \pm
\xi^\dagger Q^a\xi\right) p_\nu p_\rho p_\sigma\right\}
\nonumber \\ &&
-i\beta^\prime{\rm tr}\left\{Q^a\left(
\{{U{\cal M}+\cal M}U^{\dag},\alpha_\mu\}\mp
\{{\cal M}U+U^{\dag}{\cal M},\beta_\mu\}\right)\right\},
\label{currps}
\ee
where $Q^a=(\frac{1}{3},\frac{\lambda^1}{2},\ldots,
\frac{\lambda^8}{2})$ denote the Hermitian nonet generators.
The combination
\be
Q^{\rm e.m.}={\rm diag}
\left(\frac{2}{3},-\frac{1}{3},-\frac{1}{3}\right)
=Q^3+\frac{1}{\sqrt3}Q^8
\label{emgen}
\ee
is of special interest because it enters the computation of the 
electromagnetic properties. The associated form factors of the 
$\frac{1}{2}^+$ baryons ($B$) are defined by
\be
\langle B(\mbox{\boldmath $p$}^\prime)|V_\mu^{\rm e.m.}|
B(\mbox{\boldmath $p$})\rangle=
{\overline u}(\mbox{\boldmath $p$}^\prime)\left[
\gamma_\mu F^B_1(q^2)+
\frac{\sigma_{\mu\nu}q^\nu}{2M_B}F^B_2(q^2)\right]
u(\mbox{\boldmath $p$})\ , \qquad
q_\mu=p_\mu-p_\mu^\prime \ .
\label{defff}
\ee
Frequently it is convenient to introduce ``electric" and 
``magnetic" form factors
\be
G_E^B(q^2)=F^B_1(q^2)-\frac{q^2}{4M_B^2}F^B_2(q^2)\ , \qquad
G_M^B(q^2)=F^B_1(q^2)+F^B_2(q^2)\ .
\label{emff}
\ee

Substituting the rotating hedgehog configuration into the defining 
equation of the currents (\ref{currps}) yields for the spatial 
components of the vector current\footnote{The conventions are 
$i,j,k=1,2,3$ and $\alpha,\beta=4,\ldots,7$.}
\be
V_i^a&=&V_1(r)\epsilon_{ijk}x_jD_{ak}
+\frac{\sqrt3}{2}B(r)\epsilon_{ijk}\Omega_jx_kD_{a8}
+V_2(r)\epsilon_{ijk}x_jd_{d\alpha\beta}D_{a\alpha}\Omega_\beta
\nonumber \\ &&
+V_3(r)\epsilon_{ijk}x_jD_{88}D_{ak}
+V_4(r)\epsilon_{ijk}x_jd_{d\alpha\beta}D_{8\alpha}D_{a\beta}
+\ldots \ ,
\label{spcurrps}
\ee
where 
\be
B(r)=\frac{-1}{2\pi^2}\fp\frac{\sFt}{r^2}
\label{bdensity}
\ee
is the baryon number density (\ref{topcur}). The explicit form of 
the radial functions $V_1(r),\ldots,V_4(r)$ is given in appendix B of 
ref \cite{Pa91}. According to the quantization prescription (\ref{Rgen}) 
the angular velocities $\Omega_a$ are replaced by their expressions 
in terms of the right generators $R_a$ of $SU(3)$. Taking the Fourier 
transform of the resulting matrix elements allows one to identify the 
magnetic form factor in the Breit frame \cite{Br86,Me87}
\be
G_M^B(\mbox{\boldmath $q$}^2)&=&-8\pi M_B
\int_0^\infty \hspace{-0.2cm} r^2 dr
\frac{r}{|\mbox{\boldmath $q$}|}j_1(r|\mbox{\boldmath $q$}|)
\Bigg\{V_1(r)\langle D_{e3}\rangle_B
-\frac{1}{2\alpha^2}B(r)\langle D_{e8}R_8\rangle_B
\label{Gmps} \\ && \hspace{1cm}
-\frac{1}{\beta^2}V_2(r)\langle
d_{3\alpha\beta}D_{e\alpha}R_\beta\rangle_B
+V_3(r)\langle D_{88}D_{e3}\rangle_B
+V_4(r)\langle d_{3\alpha\beta}D_{e\alpha}D_{8\beta}\rangle_B
\Bigg\} \ .
\nonumber
\ee
Here the flavor index $e$ refers to the ``electromagnetic"
direction (\ref{emgen}). The magnetic moment corresponds to the 
magnetic form factor at zero momentum transfer $\mu_B=G_M^B(0)$.
Similarly the electric form factor is given by Fourier 
transforming the time component of the electromagnetic current
\be
G_E^B=4\pi\int_0^\infty \hspace{-0.2cm}
r^2 dr j_0(r|\mbox{\boldmath $q$}|)
\left\{\frac{\sqrt3}{2}B(r)\langle D_{e3}\rangle_B
+\frac{1}{\alpha^2}V_7(r)\langle D_{ei}R_i\rangle_B
+\frac{1}{\beta^2}V_8(r)\langle D_{e\alpha}R_\alpha\rangle_B
\right\} \hspace{-0.1cm} .
\label{Geps}
\ee
The two new radial functions $V_7(r)$ and $V_8(r)$ are listed in 
appendix B of ref \cite{Pa91} as well. Integrating $V_7$ and $V_8$ 
yields the moments of inertia, $\alpha^2$ and $\beta^2$, respectively. 
Hence the electric charges are properly normalized. It should be 
remarked that the baryon matrix elements in the space of the collective 
coordinates are computed using the exact eigenstates of (\ref{c2problem}) 
and adopting the Euler--angle representations for the $SU(3)$ 
generators \cite{We96}. The results for the magnetic 
moments and the radii
\be
r_M^2=-\frac{6}{\mu_B}\frac{d G_M^B(\mbox{\boldmath $q$}^2)}
{d\mbox{\boldmath $q$}^2}\Bigg|_{\mbox{\boldmath $q$}^2=0}
\quad {\rm and} \quad
r_E^2=-6\frac{d G_E^B(\mbox{\boldmath $q$}^2)}
{d\mbox{\boldmath $q$}^2}\Bigg|_{\mbox{\boldmath $q$}^2=0}
\label{radii}
\ee
are shown in table \ref{ta_emps}.
\begin{table}
\caption{\label{ta_emps}\small
The electromagnetic properties of the baryons compared to the 
experimental data. The predictions of the Skyrme model are taken 
from ref \protect\cite{Pa91}.}
~
\newline
\centerline{\small\smalllineskip
\begin{tabular}{c| c c | c c | c c}
 & \multicolumn{2}{c|}{$\mu_B({\rm n.m.})$} &
\multicolumn{2}{c|}{$r_M^2({\rm fm}^2)$} &
\multicolumn{2}{c}{$r_E^2({\rm fm}^2)$} \\
$B$ & $e=4.0$ & Expt. & $e=4.0$ & Expt. & $e=4.0$ & Expt. \\
\hline
$p$         & 2.03 & 2.79 & 0.43 & 0.74 & 0.59 & 0.74 \\
$n$         &-1.58 &-1.91 & 0.46 & 0.77 &-0.22 &-0.12 \\
$\Lambda$   &-0.71 &-0.61 & 0.36 & ---  &-0.08 & ---  \\
$\Sigma^+$  & 1.99 & 2.42 & 0.45 & ---  & 0.59 & ---  \\
$\Sigma^0$  & 0.60 & ---  & 0.36 & ---  &-0.02 & ---  \\
$\Sigma^-$  &-0.79 &-1.16 & 0.58 & ---  &-0.63 & ---  \\
$\Xi^0$     &-1.55 &-1.25 & 0.38 & ---  &-0.15 & ---  \\
$\Xi^-$     &-0.64 &-0.69 & 0.43 & ---  &-0.49 & ---  \\
$\Sigma^0\rightarrow\Lambda$
            &-1.39 &-1.61 & 0.48 & ---  & ---  & ---  \\
\end{tabular}
}
\end{table}
As in the two flavor model \cite{anw} the isovector part of the 
magnetic moments is underestimated while the isoscalar part is 
reasonably well reproduced. Despite the fact that the flavor 
symmetry breaking is large for the baryon wave--functions, the 
predicted magnetic moments do not strongly deviate from the 
$SU(3)$ relations \cite{Ad85}
\be
\mu_{\Sigma^+}&=&\mu_p\ ,\quad \mu_{\Sigma^0}=\frac{1}{2}
(\mu_{\Sigma^+}+\mu_{\Sigma^-})\ ,\quad 
\mu_{\Sigma^-}=\mu_{\Xi^-}\ ,
\nonumber \\
2\mu_\Lambda&=&-(\mu_{\Sigma^+}+\mu_{\Sigma^-})
=-2\mu_{\Sigma^0}=\mu_n=\mu_{\Xi^0}=
{2\over{\sqrt3}}\mu_{\Sigma^0\Lambda}\ .
\label{su3mag}
\ee
A more elaborate treatment of the flavor symmetry breaking is necessary 
in order to accommodate the experimentally observed details of breaking 
the $U$--spin symmetry which {\it e.g.} causes the approximate identity 
$\mu_{\Sigma^+}\approx\mu_p$ \cite{Sch92}.
The moderate differences between the various magnetic radii 
$r_M^2$ is a further hint that symmetry breaking effects are 
mitigated. The comparison with the available empirical data for 
the radii shows that the predictions turn out too small in 
magnitude (except for the neutron electric radius). This is a strong 
indication that essential ingredients are still missing in the 
model. In section 5 it will be explained that 
the effects, which are associated with vector meson dominance 
(VMD), will account for this deficiency. Nevertheless the overall 
picture gained for the electromagnetic properties of the 
$\frac{1}{2}^+$ and $\frac{3}{2}^+$ can at least be characterized 
as satisfactory, especially in view of the fact that 
the only free parameter of the model has been fixed beforehand.

\bigskip
\leftline{\large\bf 4.4 Effects of symmetry breaking on baryon matrix
elements}
\leftline{\large\bf~~~\,\, and strangeness in the nucleon}
\medskip

Theorists typically wish for symmetry breaking effects to be negligible
(the so--called ``spherical cow'' approximation) but Nature says
otherwise.
This is very apparent in the case of low energy strong interactions (QCD).
The Gell--Mann--Okubo mass formulae, which amount to applications of a 
Wigner--Eckart theorem for first order $\lambda_8$ type symmetry breaking,
furnish sum rules rather than a complete description. As we discussed
in section 4.2, the Skyrme model provides a non--trivial playground for 
treating symmetry breaking. The Yabu--Ando equation (\ref{c2problem})
gives an exact (within the model) wave--function for each baryon state at 
any strength of the symmetry breaking parameter $\gamma\beta^2$ ($\propto$
underlying quark masses). The physical results vary smoothly with 
$\gamma\beta^2$ (although higher quantum corrections would be expected 
to give weak non--analytic corrections).

The physical interpretation of symmetry breaking in the model may be seen
from eq (\ref{nwfexp}). The higher $SU(3)$ representation components in 
the baryon wave--function can only emerge in a quark framework by having
quark--antiquark pairs present in addition to the three ``valence'' quarks.
Clearly such effects would be difficult to treat in the non--relativistic
quark model approach. On the other hand, it should be recognized that the 
Skyrme model approach is based on a collective semi--classical treatment.

In the last few years there has been a greatly renewed interest in the
study of symmetry breaking effects for ordinary nucleons. This was 
stimulated by new experiments on polarized lepton deep inelastic 
scattering off nucleons \cite{EMC} which seem to indicate that the
pure valence quark picture of the nucleon has serious drawbacks. The
Skyrme model has the advantage of giving a simple and roughly accurate
quantitative explanation of these experiments. In detail one needs the
strangeness conserving proton matrix elements
\be
\langle P(\mbox{\boldmath $p$}^\prime)|
{\overline q}_i\gamma_\mu\gamma_5 q_i|
P(\mbox{\boldmath $p$})\rangle=
{\overline u}(\mbox{\boldmath $p$}^\prime)\left[
\gamma_\mu H_i(q^2)+
\frac{q_\mu}{2M_p}\tilde H_i(q^2)\right]\gamma_5
u(\mbox{\boldmath $p$})
\label{paxmat}
\ee
for this discussion. Of related interest are the flavor changing matrix 
elements 
\be
\langle B^\prime(\mbox{\boldmath $p$}^\prime)|V_\mu^\alpha|
B(\mbox{\boldmath $p$})\rangle & = &
{\overline u}(\mbox{\boldmath $p$}^\prime)\left[
\gamma_\mu g_V(q^2)+\ldots\right]u(\mbox{\boldmath $p$})\ ,
\nonumber \\
\langle B^\prime(\mbox{\boldmath $p$}^\prime)|A_\mu^\alpha|
B(\mbox{\boldmath $p$})\rangle & = &
{\overline u}(\mbox{\boldmath $p$}^\prime)\left[
\gamma_\mu\gamma_5 g_A(q^2)+\ldots\right]
u(\mbox{\boldmath $p$})\ .
\label{semlepmat}
\ee
between different baryons ($B^\prime,B$). Here we have omitted 
contributions proportional to the momentum transfer $q_\mu$.

Knowledge of the $g_A(B,B^\prime)$ and $g_V(B,B^\prime)$ are 
crucial for the theory of baryon semi--leptonic decays. First
let us consider the calculation of the axial vector matrix
elements $g_A(B,B^\prime)$. Our main interest in this brief 
discussion will be to examine the effects of symmetry breaking.
The leading order term (in $1/N_C$) of the spatial components of 
the axial current is straightforwardly obtained to be
\be
\int d^3r A_i^a = {\cal C}D_{ai}(A) \ ,
\label{aiagen}
\ee
where $A(t)$ is the collective coordinate matrix.
The constant ${\cal C}$ denotes an integral over the chiral angle.
We refer the interested reader to refs \cite{Pa90,Pa91} for the 
explicit expression. Then,
\be
g_A^a(B^\prime,B)=
{\cal C}\langle B^\prime | D_{a3} | B \rangle \ .
\label{calgaa}
\ee
The flavor index $a$ has to be chosen according to whether strangeness 
conserving ($a=1,2,3,8$) or strangeness changing ($a=4,\ldots,7$) 
processes are being considered. The corresponding result for the axial 
charge of the nucleon $g_A=g_A^{1+i2}(p,n)$, as measured in 
neutron beta--decay, is predicted too low in many soliton models.
This problem is already encountered in the two 
flavor model and gets worse in $SU(3)$ as the Clebsch--Gordon 
coefficient associated with $D_{1+i2\ 3}$ changes by a factor 
of 7/10. As symmetry breaking is increased the $SU(3)$ prediction 
for $g_A$ becomes larger \cite{Pa89}
\be
g_A(SU(3))=\frac{7}{10}\left[1
+0.0514\gamma\beta^2+\ldots\right]g_A(SU(2)) \ .
\label{expga}
\ee
Actually the exact treatment shows that with increasing symmetry 
breaking the two flavor result is approached, although only slowly.
Taking everything together, including subleading terms in 
(\ref{calgaa}), finally gives $g_A=0.98$ for $e=4.0$ \cite{Pa91}
which is about 4/5 of the experimental value 
$g_A({\rm expt.})=1.26$. 

We may understand the tendency to approach the $SU(2)$ limit for 
large $\gamma\beta^2$ as follows. In the small $SU(3)$ breaking case,
there is just a small extra ``cost'' for producing an $\bar{s}s$
pair rather then a $\bar{u}u$ or $\bar{d}d$ pair. As $\gamma\beta^2$ 
gets larger it is more expensive to make an $\bar{s}s$ pair and 
eventually $\bar{s}s$ pairs should be absent from the nucleon 
wave--function, recovering the $SU(2)$ picture.

Returning to the general case one should first note that flavor symmetry 
relates the octet axial current matrix elements between various 
baryons. Conventionally they are expressed using $SU(3)$ covariance in 
terms of two unknown constants (or reduced matrix elements) $F$ and $D$. 
One has to use models to determine these constants. In the flavor symmetric 
Skyrme model one finds \cite{Ad85} $D/F=9/5$ and $D+F=7{\cal C}/15=g_A$. 
In table \ref{ta_hyperon} the flavor symmetric dependences of the axial
matrix elements on $F$ and $D$ are displayed. 
\begin{table}
\caption{\label{ta_hyperon}\small
The matrix elements of the axial--vector current (\protect\ref{calgaa})
between different baryon states in the flavor symmetric limit.
Displayed are both the strangeness conserving (a) and strangeness 
changing (b) processes. The first column gives the relevant 
flavor component of the axial current.}
~
\newline
\centerline{\small\smalllineskip
\begin{tabular}{c| c c c c}
& \multicolumn{4}{c}{(a)} \\
& $n\rightarrow p$ & $\Sigma^-\rightarrow\Lambda$ &
$\Sigma^-\rightarrow\Sigma^0$ & $\Xi^-\rightarrow\Xi^0$ \\
$A^{\pi^-}$ & $F+D$ & $\frac{2}{\sqrt6}D$ & $\sqrt2 F$ & $D-F$ \\
\hline
& \multicolumn{4}{c}{(b)} \\
& $\Lambda\rightarrow p$ & $\Sigma^-\rightarrow n$ &
$\Xi^-\rightarrow\Lambda$ & $\Xi^-\rightarrow\Sigma^0$ \\
$A^{K^-}$ & $\frac{1}{\sqrt6}(3F+D)$ & $D-F$ & 
$\frac{1}{\sqrt6}(3F-D)$ & $\frac{1}{\sqrt2}(F+D)$ \\
\end{tabular}}
\end{table}
As one departs from the flavor symmetric case the baryon 
wave--functions acquire admixture from higher dimensional $SU(3)$ 
representations making the $SU(3)$ covariant parameterization in 
terms of $F$ and $D$ inadequate. In the presence of $SU(3)$ symmetry
breaking we must, without further assumptions, parameterize each decay
amplitude separately. It is still reasonable to maintain the isospin
invariance relations.

\begin{figure}[t]
\parbox[l]{10.5cm}{
\centerline{\hskip -1.5cm
\epsfig{figure=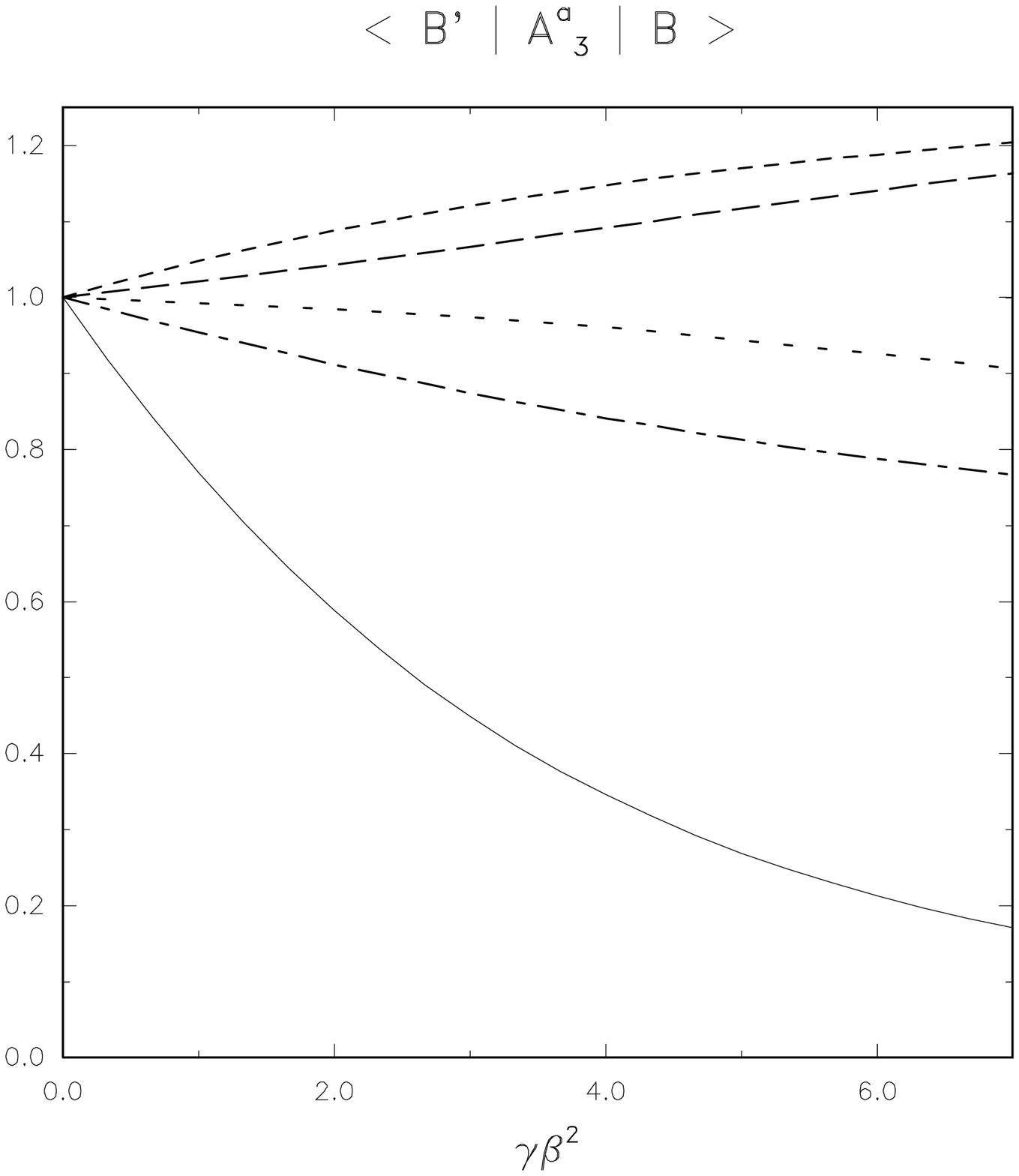,height=6.5cm,width=10.0cm}}}~
\parbox[r]{6.0cm}{\vskip-1.0cm
\caption{\label{fi_axial}
The variation of axial vector matrix elements with the effective 
symmetry breaking parameter~${\Ts \gamma\beta^2}$.
\newline
Full line:
${\Ts \langle p|{\overline{s}}\gamma_3\gamma_5s| p\rangle}$;~
dashed dotted line:
${\Ts \langle p|{\overline{u}}\gamma_3\gamma_5s| \Lambda\rangle}$;~
dotted line:
${\Ts \langle n|{\overline{u}}\gamma_3\gamma_5s|
\Sigma^-\rangle}$;~
long dashed line:
${\Ts \langle\Lambda|{\overline{u}}\gamma_3\gamma_5s|
\Xi^-\rangle}$;~
dashed line:
${\Ts \langle p|{\overline{u}}\gamma_3\gamma_5d|
n\rangle}$.
These matrix elements, which are taken from refs 
\protect\cite{Pa90} and \protect\cite{Pa89a}, are normalized to the 
flavor symmetric values.}}
\end{figure}

As an example of the perturbative corrections consider the axial 
$\Lambda\to p$ transition in the Cabibbo scheme \cite{Ca63} for 
semi--leptonic hyperon decays. The analog of (\ref{nwfexp}) for the 
$\Lambda$ hyperon is
\be
|\Lambda\rangle=|\Lambda, {\bf 8}\rangle 
+\frac{3}{50}\gamma\beta^2|\Lambda, {\bf 27}\rangle+\ldots \ .
\label{lamwfexp}
\ee
Noting that the $D$--functions mix different $SU(3)$ representations, 
we get
\be
\langle p \uparrow | D_{K^-3} |\Lambda \uparrow\rangle =
\frac{2}{5\sqrt3}-\frac{7\sqrt3}{1125}\gamma\beta^2+\ldots \ , 
\qquad
D_{K^-3}=\frac{1}{\sqrt2}\left(D_{43}-iD_{53}\right) \ .
\label{lampexp}
\ee 
Of course, this expansion just provides a first approximation to 
the symmetry breaking dependence of the Cabibbo matrix elements. 
Using the exact treatment initiated by Yabu and Ando \cite{Ya88} 
this dependence can be computed numerically as shown in figure
\ref{fi_axial} for some processes of interest \cite{Pa90,Pa89a}. 
Those results are normalized to the $SU(3)$ symmetric values in 
table \ref{ta_hyperon} to illustrate that the matrix elements vary 
in different ways with symmetry breaking.
In this figure also the variation of the nucleon matrix element of 
the flavor conserving axial current 
$H_3(0)=\langle N|{\overline s}\gamma_3\gamma_5s|N\rangle$ is 
displayed. Obviously $H_3(0)$ decreases very rapidly with increasing
symmetry breaking. This is easily visualized, as mentioned above, as
a reflection of the increased cost of making extra $\bar{s}s$ pairs 
in the nucleon wave--function as $\gamma\beta^2$ increases. On the 
contrary the Cabibbo matrix elements exhibit only a moderate dependence 
on $\gamma\beta^2$. It is this different behavior of the matrix elements 
that makes the application of exact flavor symmetry to the analyses of 
the EMC--SLAC--SMC experiments suspicious. Stated otherwise, 
the strange quark contribution to the nucleon matrix element of the 
axial singlet current (loosely ``proton spin'') may be decreased 
significantly as a consequence of symmetry breaking without 
contradicting the successful Cabibbo scheme for the semi--leptonic 
decays of the hyperons. 

As an interesting contrast to the axial matrix elements, consider 
the evaluation of the vector matrix elements $g_V(B,B^\prime)$ needed
for the hyperon semi--leptonic decays. The dominant contribution is 
given by the matrix elements of the $SU(3)$ flavor generators
\be
g_V^a(B^\prime,B)=\langle B^\prime | L_a | B \rangle \ .
\label{calgva}
\ee
For example if we sandwich the generators $L_{K^-}$ between the perturbative
$\Lambda$ and $p$ states given in (\ref{lamwfexp}) and (\ref{nwfexp}) 
and recognize that group generators can only connect states belonging 
to the same irreducible representation, we see that symmetry breaking
corrections start out as $\left(\gamma\beta^2\right)^2$ rather than 
$\gamma\beta^2$. This is just a demonstration of the 
Ademollo--Gatto theorem \cite{Ad64}, which ``protects'' the vector
matrix elements against small symmetry breaking corrections. Since
$\gamma\beta^2$ is large the numerical validity of this result is 
questionable. However, the exact Yabu--Ando scheme does confirm 
\cite{Pa90} that vector matrix elements suffer at most, 10\% deviation 
from the symmetric values, even for large symmetry breaking, {\it e.g.}
$\gamma\beta^2\approx7$. 

A reduction of strangeness in the nucleon with increasing $\gamma\beta^2$
is also predicted for the scalar strange content fraction of the proton
\be
X_s=\frac{\langle p|{\overline s}s|p\rangle
-\langle0|{\overline s}s|0\rangle}
{\langle p|{\overline u}u+{\overline d}d+{\overline s}s|p\rangle-
\langle 0|{\overline u}u+{\overline d}d+{\overline s}s|0\rangle} \ .
\label{defxs}
\ee
Here the state $|0\rangle$ refers to the soliton being absent.
Models of quark flavor dynamics, as {\it e.g} the one of 
Nambu--Jona--Lasinio \cite{NJL}, indicate that matrix elements
of quark bilinears $\bar{q}\lambda_a q$ may be taken as 
proportional to the matrix elements of ${\rm tr}\, 
\left[\lambda_a\left(U+U^\dagger-2\right)\right]$. Then we 
straightforwardly get
\be
X_s=\frac{1}{3}\langle p |1-D_{88}| p \rangle
\approx\frac{7}{30}-\frac{43}{2250}\gamma\beta^2+\ldots \ .
\label{xsskyrme}
\ee
In this case, however, the deviation from the flavor symmetric
result \cite{Do86} ($X_s=7/30$) is considerably mitigated \cite{Ya89}
as compared to the variation of $H_s$ defined in (\ref{paxmat}). 
The symmetry breaking has to be as large as 
$\gamma\beta^2\approx4.5$ to obtain a reduction of the
order of 50\%. In the case of $H_s$ this was already achieved for
$\gamma\beta^2\approx2.5$. In any event, the additional
quark--antiquark excitations in the nucleon, which are
parametrized by the admixture of higher dimensional $SU(3)$
representations (\ref{nwfexp}), clearly tend to cancel the
virtual strange quarks of the pure octet nucleon.

The three flavor Skyrme model under present consideration provides
a convenient way to study the nucleon matrix elements of the vector
current ${\bar s}\gamma_{\mu}s$ . These are theoretically interesting
because they would vanish in a pure valence quark model of the nucleon
and so test finer details of nucleon structure. They are experimentally
interesting because they can be extracted from measurements of the
parity violating asymmetry in the elastic scattering of polarized
electrons from the proton. The precise form factors needed are defined
by
\be
\langle P(\mbox{\boldmath $p$}^\prime)|
{\overline q}_s\gamma_\mu q_s|
P(\mbox{\boldmath $p$})\rangle=
{\overline u}(\mbox{\boldmath $p$}^\prime)\left[
\gamma_\mu F_s(q^2)+
\frac{\sigma_{\mu\nu}q^\nu}{2M_p}\tilde F_s(q^2)\right]
u(\mbox{\boldmath $p$})\ .
\label{pvecmat}
\ee
These form factors are currently under intensive experimental 
investigation, {\it cf.} refs \cite{samhap} and have been 
estimated in various models. The models range from vector--meson--pole 
fits \cite{Jaf89} of dispersion relations
\cite{Ho74} through vector meson dominance approaches \cite{Pa91}
and kaon--loop calculations with \cite{Mu93} and without \cite{Fo94}
vector meson dominance contributions to soliton model calculations
\cite{Pa91,Pa92,We95a}. The numerical results for the strange
magnetic moment $\mu_S=\tilde F_s(0)\approx
-0.31\pm0.09\ \ldots \ 0.25$ are quite diverse. The predictions
for the strange charge radius $r_S^2=-6dF_s(q^2)/dq^2|_{q=0}$
are almost equally scattered $r_S^2\approx -0.20\ \ldots \
0.14 {\rm fm}^2$.

 In order to evaluate these form factors
in the three flavor Skyrme model one requires the matrix
elements of the ``strange" combination
\be
Q^s=\frac{1}{3}\ID-\frac{1}{\sqrt3}\lambda_8=
Q^0-\frac{2}{\sqrt3}Q^8
\label{strgen}
\ee
rather than the electromagnetic one (\ref{emgen}) between proton states. 
Using the same value $e=4.0$ as used consistently
for the three flavor pseudoscalar model yields the predictions
\be
\mu_S=-0.13 {\rm n.m.} \ , \qquad
r_S^2=-0.10 {\rm fm}^2 \ .
\label{stresps}
\ee
Here n.m. stands for nuclear magnetons.
It should be stressed that these results are obtained within
the Yabu--Ando approach, {\it i.e.} the proton wave--function
contains sizable admixture of higher dimensional representations.
If a pure octet wave--function were employed to compute the
matrix elements of the collective operators the strange magnetic
moment would have been $\mu_S=-0.33$. The proper inclusion of
symmetry breaking into the nucleon wave--function is again seen to 
reduce the effect of the strange degrees of freedom in the nucleon. 
We already discussed above that as the strange quarks within the nucleon 
become more massive (effect of symmetry breaking) their excitation becomes
less likely.

In the next few years a great deal of new experimental information
on the form factors $F_s$ and $\tilde F_s$ should become available.
This would enable more accurate comparison with (for a given
effective meson Lagrangian) the essentially parameter free predictions
of the soliton theory.

\bigskip
\stepcounter{chapter}
\leftline{\Large \bf 5. The nucleon as a vector meson soliton}
\medskip

According to the modern view of the Skyrme model approach we
should start from the ``full" effective Lagrangian which contains
mesons of all spins. The practical criterion on which particles to include
is to find an effective Lagrangian which does a good job of
explaining the low energy experimental data in the meson sector. On
these grounds it is evident that the vector mesons should be included in
the effective Lagrangian. In this section we give a brief sketch
of the soliton sector of the Lagrangian of pseudoscalars and
vectors (see section 3.3) and note that it leads to significant 
improvements of many predictions. In particular it is crucial for
discussing the so--called {\it proton spin puzzle}.

\bigskip
\leftline{\large\bf 5.1 Generalized soliton ansatz and profile functions}
\medskip

\begin{figure}[ht]
\parbox[l]{11cm}{
\centerline{
\epsfig{figure=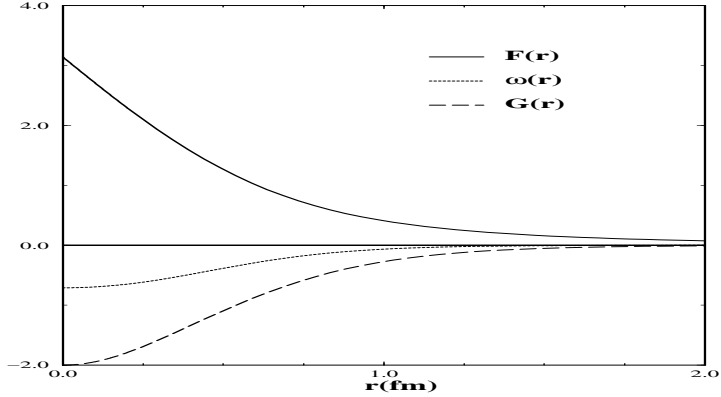,height=5.0cm,width=10.0cm}
}}~~
\parbox[r]{5.5cm}{
\caption{\label{fig_vm}The profile functions which minimize
the classical vector meson energy functional~(\protect\ref{mcl})
for the parameters $g=5.85$, $\tilde{h}=0.4$, $\tilde{g}_{VV\Phi}=1.9$
and $\kappa=1.0$. $\omega(r)$ is measured in units of $m_\rho$.}}
\end{figure}

As a first step we construct the soliton of the Lagrangian 
defined in section 3.3. The generalization of the hedgehog {\it ansatz} 
(\ref{hedgehog}) to the vector meson model requires the time component 
of the $\omega$ field and the space components of the $\rho$ field to
be different from zero.  Parity and grand spin symmetry\footnote{The
grand spin {\bf J} +{\bf I} is the characteristic invariance of the
hedgehog ansatz.}  allow for three radial functions
\be
\xi_\pi=
{\rm exp}\left(\frac{i}{2}\hat{\mbox{\boldmath $r$}}\cdot
{\mbox{\boldmath $\tau$}} F(r)\right)\ ,\quad
\omega_0={{\omega(r)}\over{2g}}\ ,\quad
\rho_i^a={{G(r)}\over{gr}}\epsilon_{ija}\hat r_j\ .
\label{vmhedgehog}
\ee
Substituting these {\it ans\"atze} into the action described in 
section 3.3 yields the classical mass, 
\be
E&=&4\pi\int dr\Big[{\ft\over 2}(\fpt r^2+2\sFt)
-{r^2\over{2g^2}}(\wpt+m_{\rho}^2\wo^2)
+{1\over{g^2}}[\gpt+{{G^2}\over{2r^2}}(G+2)^2]
\nonumber \\
&&+{{m_\rho^2}\over{g^2}}(1+G-\cF)^2+{\go\over g}\fp \wo \sFt-
{{2\gw}\over g}\gp\wo \sF
\nonumber \\
&&+{\gt\over g}\fp\wo G(G+2)+{1\over g}(\gw+\gt)\fp\wo
[1-2(G+1)\cF+\cFt]
\nonumber \\
&&+(1-\cF)\Big\{4\delta^\prime r^2+2(2\bep-{\alp\over g^2})
(\fpt r^2 +2\sFt)
\nonumber \\
&&-{{2\alp}\over{g^2}}\big[\wo^2r^2-2(G+1-\cF)^2-4(1+\cF)(1+G-\cF)
\big]\Big\}\Big].
\label{mcl}
\ee
Application of the variational principle to this
functional yields second order coupled non--linear differential
equations for the radial functions $F(r)$, $\omega(r)$ and $G(r)$. The
boundary conditions for the chiral angle $F(r)=\pi$ and $F(\infty)=0$,
which correspond to unit baryon number, also determine the boundary
conditions of the  vector meson profiles via the differential equations
and the requirement of finite energy.
For example we find
$G(0)=-2$. A typical set of resulting profile functions
is shown in figure \ref{fig_vm}.

As in the pseudoscalar model we have to generate states with good
spin and isospin from this classical field configuration. To start
with, one introduces collective coordinates (\ref{colcor1}) for all 
fields which have non--vanishing spin or isospin. However, an additional 
complication arises because there are vector meson field components which 
vanish classically but get excited by the collective rotation. In the 
two flavor case the appropriate {\it ansatz} for these excitations reads
\be
\mbox{\boldmath $\rho$}_0
=\frac{1}{2g}A(t)\left[\xi_1(r)\mbox{\boldmath $\Omega$}
+\xi_2(r)(\hat{\mbox{\boldmath $r$}}\cdot\mbox{\boldmath $\Omega$})
\hat{\mbox{\boldmath $r$}}\right]A^\dagger(t)\ , \quad
\omega_i=\frac{\Phi(r)}{2g}\epsilon_{ijk}\Omega_j\hat r_k\ ,
\label{vmind1}
\ee
where the angular velocity of the rotating soliton, $\Omega_i$ is defined
in (\ref{colcor3}).
The three radial functions $\xi_1,\xi_2$ and $\Phi$ are not the only 
ones which get excited. As these radial functions are non--zero they 
provide sources for the non--strange component of the iso--singlet 
pseudoscalar field via the $\epsilon$--terms (\ref{anom2}). In the two
flavor formulation the appropriate {\it ansatz} which takes into account 
the pseudoscalar nature of the $\eta$ field reads
\be
U(\vecbf{r},t)={\rm e}^{i\eta_T(\vecbfs{r})}
A(t)U_0(\vecbf{r})A^\dagger(t) \quad {\rm with} \quad
\eta_T(\vecbf{r})=\frac{1}{f_\pi}\eta(r)\hat{\vecbf{r}}\cdot
\vecbf{\Omega}\, .
\label{etaex}
\ee
As we will observe shortly, the excitation of this $\eta$ field plays 
a decisive role in the context of the {\it proton spin puzzle}. The 
additional radial functions are determined from extremizing the moment 
of inertia for rotations in coordinate space,
\be
\alpha^2&=&
{8\pi\over 3}\int dr \Bigg\{\ft r^2
\sFt-{4\over{g^2}}(\phi^{\prime2}+2{\phi^2\over r^2}+m_\rho^2\phi^2)
+{{m_\rho^2}\over{2g^2}}r^2\left[(\xi_1+\xi_2)^2+2(\xi_1-1+\cF)^2\right]
\nonumber \\*
&&+{1\over{2g^2}}\left[(3\xi_1^{\prime2} +2\xi_1^\prime\xi_2^\prime+
\xi_2^{\prime2})r^2+2(G^2+2G+2)\xi_2^2+4G^2(\xi_1^2+\xi_1\xi_2
-2\xi_1-\xi_2+1)\right]
\nonumber \\
&&+{4\over g}\go \phi\fp \sFt
+{4\over g}\gt\phi\fp\left[(G-\xi_1)(1-\cF)+(1-\cF)^2-G\xi_1\right]
\nonumber \\
&&+{{2\gw}\over g}\Big\{\phi^\prime \sF(G-\xi_1+2-2\cF)
+\phi \sF(\xi_1^\prime-G^\prime)
\nonumber \\
&&+\phi\fp\left[
2+2\sFt+(\xi_1-G-2)\cF-2(\xi_1+\xi_2)\right]\Big\}
\nonumber \\
&&\qquad
-{1\over2}\left[\eta^{\prime2}r^2+2\eta^2
+ m_\eta^2r^2\eta^2\right]
+{{\gw g}\over{2f_\pi}}\left[\eta(\phi\wop -\wo\phi^\prime)
-\eta^\prime\phi\wo\right]
\nonumber \\
&&-{\go\over{3gf_\pi}}\left[\eta^\prime
(\xi_1+\xi_2)\sFt+2\eta\fp(G+\xi_1)\sF\right]
-{{3\gt}\over{gf_\pi}}\eta^\prime(G+1-\cF)^2(\xi_1+\xi_2)
\nonumber \\*
&&-{\gw\over{gf_\pi}}\left\{\eta^\prime\left[(G+\xi_1)G+(\xi_1+\xi_2)
[(1-\cF)^2-2G\cF]\right]+\eta(G\xi_1^\prime-\gp\xi_1)\right\}\Bigg\}\, ,
\label{al2vm}
\ee
together with suitable boundary conditions. In eq (\ref{al2vm}) we
have not displayed the explicit contributions from the symmetry
breakers (which are in fact small). We
will mostly limit 
the present discussion to the two flavor case. In the case of three
flavor vector meson models the situation is even more complicated
as also $K^*$ type fields get excited. Also there will be additional 
symmetry breakers on the level of the collective Lagrangian which are 
of the form $\sum_{i=1}^3 D_{8i}\Omega_i$ and stem from terms which are 
linear in the time derivative. They can straightforwardly be implemented 
in the collective quantization approach. Here we will omit details but
rather refer the reader to the literature \cite{Pa92,We96}. The
general pattern for computing baryon properties is essentially the same
as that discussed for the Lagrangian of only pseudoscalars in section 4.

\bigskip
\leftline{\large\bf 5.2 Axial singlet current and proton spin
puzzle}
\medskip

Notice that in (\ref{al2vm}) we included by hand a mass term for the
rotationally excited profile $\eta(r)$ of a pseudoscalar isosinglet
field. Actually the existence of such a term has not yet been justified.
Before proceeding we must do so since the term turns out to be very
important.

In section 3.1 we mentioned that the QCD axial singlet current
\be
J^{(0)}_{5,\mu} = {\bar u}\gamma_\mu\gamma_5u + {\bar d}\gamma_\mu
\gamma_5d + {\bar s}\gamma_\mu\gamma_5s,
\label{asc}
\ee
is not conserved even for zero quark masses: $\partial^\mu J^{(0)}_{5,\mu}
=G$, where the $U_A(1)$ anomaly $G$ is proportional to the product of
the QCD field strength tensor $F^a_{\mu\nu}$ and its dual. In order to
mock up this non--conservation equation at the effective Lagrangian level
\cite{U1A} we may add the terms
\be
\frac{c}{2}G^2 + 
\frac{iG}{12}{\rm ln}
\left(\frac{{\rm det}\, U}{{\rm det}\, U^{\dagger}}\right),
\label{ant}
\ee
where $G$ is now considered a composite glueball field which ``dominates"
the $U_A(1)$ anomaly. Here we assumed three light flavors and also that
the strong CP violation parameter $\theta$ is zero. Furthermore it
is necessary that, except for the terms representing quark mass symmetry
breaking, all the other terms in the effective Lagrangian be invariant
under $U_A(1)$. The parameter c above is determined by
\be
m^2_{\eta_0} \approx \frac{1}{6cf^2_\pi} ,
\label{etaprime}
\ee
in the approximation where the quark mass terms are neglected. $\eta_0$
is the $SU(3)$ singlet pseudoscalar field as in (\ref{su3chif}). This
equation arises after noting that G is like an auxiliary field and may be
integrated out: $G = \eta_0/(\sqrt{6}cf_\pi)$ . In the effective
Lagrangian the realization of the axial singlet current, obtained
by a Noether variation, is
\be
J_{5,\mu}^{(0)}=\sqrt{6}f_\pi\partial_\mu \eta_0 + 
\tilde{J}_{5,\mu}^{(0)}\, .
\label{decomp}
\ee
Here the first term is the contribution from the pseudoscalar field and
the second term is due to the addition of vector fields.
$\tilde{J}_{5,\mu}^{(0)}$ has a complicated structure but, in
particular, contains {\it non--derivative} terms like
$\epsilon^{\mu\nu\alpha\beta}{\rm tr}(\rho_\nu\rho_\alpha\rho_\beta)$.
Using this decomposition we may write the equation of motion for the
$\eta_0$ field as
\be
(\partial^2+m_{\eta_0^2})\eta_0 = 
\frac{1}{\sqrt{6}f_\pi}\partial^\mu\tilde{J}_{5,\mu}^{(0)}\, ,
\label{KGE}
\ee
which shows that the vector meson contribution to the axial singlet
current may act as a source for a non--trivial excitation
associated with the $\eta_0$ field in the soliton sector.

Now the form factors for the proton matrix elements of the axial singlet
current are obviously just the sums of the three separate form factors
introduced in (\ref{paxmat}):
\be
H(q^2) = H_u(q^2) + H_d(q^2) +H_s(q^2) \quad {\rm and} \quad
\tilde H(q^2) = \tilde H_u(q^2) + \tilde H_d(q^2) + \tilde H_s(q^2).
\label{Htot}
\ee
If the vector mesons are not present, eq (\ref{decomp}) shows that the
operator for the axial singlet current must be (even in the soliton
sector) a pure derivative. This means that, regardless of the details
of the calculation, the matrix element for the sum of the three terms in
(\ref{paxmat}) must be proportional to the momentum transfer $q_\mu$ .
Thus $\tilde H(q^2) $ is non--zero and $ H(q^2) = 0 $. From the theory
of Dirac particles we recognize that the quantity $ H(0) $ has the
interpretation
of twice the quark spin part of the proton's angular momentum. We
see that the Skyrme model of only pseudoscalars predicts that the
expectation value of the net quark spin operator vanishes; the
total angular momentum (1/2) of the proton must involve,
at a fundamental level, the rotational
and gluonic pieces! Note that the above argument for $ H(0) = 0 $
with the Lagrangian of only pseudoscalars continues to hold even if
symmetry breaking contributions are taken into account \cite{Pa89a}.

The situation is a little different when vector mesons are included in
the effective Lagrangian. Since $ \tilde J_{5,\mu}^{(0)}$ has pieces 
which are not pure derivatives it then is possible to obtain
$ H(0) \ne 0 $ . A convenient parameterization for this calculation in the
effective Lagrangian model is
\be
\langle P (\vecbf{p}^\prime) | \tilde{J}_{5,i}^{(0)} 
|P  (\vecbf{p})\rangle= H(q^2) \langle 2 J_i \rangle\ .
\label{axsing1}
\ee
 Once all the radial functions 
have been determined as before from the appropriate variational
principles, it 
is straightforward to compute $H(0)$ from eq (\ref{axsing1}). One
only has to recall that under the collective coordinate quantization
the angular momentum operator is given by
$\vecbf{J}=\vecbf{\Omega}/\alpha^2$.
The numerical results for a variety of allowed parameters ({\it cf.}
discussion after eq (\ref{anom2})) are displayed in table \ref{ta_h0}.
\begin{table}[ht]
\caption{\label{ta_h0}\small
Predictions for the matrix element of the proton axial singlet current 
for various allowed sets of parameters in the vector meson model. 
Results are taken from ref \protect\cite{Jo90}.}
~
\newline
\centerline{\small\smalllineskip
\begin{tabular}{c | c c c c c c c}
$\tilde{h}$ & 0.4 & 0.4 & 0.4 & 0.7 & 0.5 & 0.2 & 0.1 \\
$\tilde{g}_{VV\Phi}$ & 1.9 & 1.9 & 1.9 & 2.2 & 2.0 & 1.7 & 1.5 \\
$\kappa$ & 0.0 & 0.5 & 1.0 & 0.0 & 0.0 & 0.0 & 0.0 \\
\hline
$H(0)$ & 0.34 & 0.33 & 0.30 & 0.29 & 0.34 & 0.32 & 0.28 \\
\end{tabular}}
\end{table}
Surprisingly the predictions of the vector meson model for 
$H(0)$ are very robust against possible changes of 
the parameters of the model. 

    Even though we get a non--zero value for $ H(0) $ in the 
vector meson model it is still small compared to $ H(0) = 1 $, the
expectation from the simple non--relativistic quark model.
Qualitatively the soliton model results with and without vectors
are similar. Since one has a natural prejudice that the quark model
results should be roughly correct, this would at first seem to
be a serious defect of the soliton approach to nucleon
properties.

Of course one can only make an accurate judgment on the
matter by appealing to experiment. $H(0) $ can be found from
eq (\ref{Htot}) if we can experimentally obtain separately
$ H_u(0), H_d(0) $ and $ H_s(0) $. The linear combination
\be
H_u(0) - H_d(0) = g_A = 1.257 ,
\label{betade}
\ee
is reliably obtained by an isotopic spin rotation of the axial
form factor describing neutron beta--decay. Similarly the estimate
for the ``eighth" octet component
\be
H_u(0) + H_d(0) - 2H_s(0) \approx 0.575 \pm 0.016 ,
\label{hypde}
\ee
may be gotten from an $SU(3)$ flavor rotation of the data on hyperon
beta--decay experiments. Clearly one more linear combination
is needed in order to disentangle the individual $ H_i(0) $
and that situation existed for many years. About ten years ago
different experimental groups (EMC, SLAC, SMC) used polarized lepton
beams to probe the structure of nucleons. The deep inelastic scattering
data \cite{EMC} were used to extract the combination
\be
4H_u(0) + H_d(0) + H_s(0) ,
\label{deepin}
\ee
in which the axial current form factor for each quark is weighted
proportionally to the square of the quark electric charge.
Combining these data 
resulted in $H(0)\approx 0.3$ \cite{Br88}. The experimental results were 
later on confirmed \cite{SMC,SLAC,SLAC98}. However, it turns out
that the theoretical extraction of $H(0)$ is quite complicated as 
it involves a careful treatment of perturbative QCD corrections.
The value \cite{El96}
\be
H(0)=0.27\pm0.04
\label{elkr}
\ee
is nowadays considered correct. At the time this low value was considered
hard to understand and the situation was called the {\it proton spin
puzzle}. We have just seen that the soliton approach does however
provide a simple explanation of such a low value.

Clearly the prediction of the vector meson treatment described in
Table 5.1, yielding H(0) about 0.30, is in good agreement with the data.
From this we learn two things. First, the simplest quark model does
not give a good description of the spin structure of the nucleon.
Second, the soliton approach based on an effective Lagrangian
including vector mesons markedly improves the qualitatively reasonable
predictions of the soliton treatment based on a pseudoscalars only 
effective Lagrangian. A physical interpretation of the latter statement is
that the pseudoscalars only Lagrangian mainly probes the ``pion cloud''
of the nucleon while the vector Lagrangian probes a little more deeply.

For completeness we remark on a possible caveat. The estimate of
(\ref{hypde}) is based on the use of exact $SU(3)$ symmetry. However in Fig
4.1 of section 4.4 we showed that precisely this current matrix element is
expected to exhibit stronger suppression than others due to $SU(3)$ symmetry
breaking. Nevertheless it turns out that \cite{Jo90} the numerical
evaluation
of $H(0)$ is not very sensitive to this feature. This is to be contrasted 
with the behavior of $H_s(0)$, which decreases rapidly with symmetry
breaking, {\it cf.} section 4.4.

\bigskip
\leftline{\large\bf 5.3 Other improvements with vector mesons}
\nopagebreak
\medskip
\nopagebreak
The famous problem of explaining the neutron--proton mass difference
is another one which requires the addition of vector mesons to the
effective Lagrangian in order to obtain a satisfactory solution
in the nucleon--as--soliton picture. It is known that the 
electromagnetic ({\it i.e.} one photon loop) contribution has the 
wrong sign. After correcting for the electromagnetic interaction the 
remaining ``strong" part of the neutron--proton mass--difference should 
be $(M_n-M_p)_{\rm strong}\approx(2.0\pm0.3){\rm MeV}$ \cite{Ga84}. At
the quark level this arises from the down quark--up quark mass difference,
controlled by the parameter $y$ in eqs (\ref{lsb}) and (\ref{qmr}). 
Information on $y$ can be most easily gained by analyzing the $K^+$--$K^0$
mass--difference, yielding $y\approx(-0.4...-0.2)$ \cite{Sch93}. To 
understand the problem it is helpful to consider the contribution of 
the (presumably dominating) $\delta^\prime$--type symmetry breaker to 
the neutron--proton mass--difference. Since the $d$--$u$ quark mass 
difference clearly exists with only two flavors it is interesting to 
first consider the problem at this level. Then the relevant piece of 
the $\delta^\prime$ term is proportional to
\be
{\rm tr}\left[\tau_3\left(U+U^\dagger\right)\right] .
\label{2fl}
\ee
Using the ansatz (\ref{etaex}) we see that 
$U = {\rm exp} (i\eta_T)[{\rm cos}(\psi)+ 
i{\vecbf n}\cdot{\vecbf \tau}{\rm sin}(\psi)]$, 
where $\psi$ is some angle. Then (\ref{2fl}) is proportional to 
${\rm sin}(\eta_T)$. In other words the contribution vanishes unless 
the field $\eta_T$ gets excited due to the collective rotation (or 
any kind of symmetry breaking). Now (\ref{KGE}) together with 
(\ref{decomp}) shows that this will not happen if only pseudoscalars 
are present in the effective Lagrangian; the vector meson contribution 
${\tilde J}_{5,\mu}^{(0)}$ must also be present. This is analogous to
the situation concerning the {\it proton spin puzzle}. The contribution 
of the $\delta^\prime$ term turns out to be
\be
(M_n-M_p)_{\rm strong}=-\frac{2y\delta^\prime}{3\alpha^2}
\int d^3r \ {\rm sin}F(r)\eta(r)+\ldots \, .
\label{mnmp}
\ee
Using the full two--flavor vector meson result for
$\tilde{J}^{(0)}_{5,\mu}$ which was already employed to compute $H(0)$
yields \cite{Ja89}
\be
(M_n-M_p)_{\rm strong}\approx1.4\, {\rm MeV}
\label{mnvm}
\ee
which, not surprisingly, turns out to be about as robust against changes 
of the parameters as is $H(0)$. This prediction is still somewhat 
too small when compared to the empirical value. However, it turns out
that the missing $\sim 0.5{\rm MeV}$ can be attributed to three flavor 
effects as matrix elements of $D_{38}$ are non--vanishing\footnote{It 
should be remarked that $\langle p |D_{38}|p \rangle$ quickly approaches 
zero as $SU(3)$ symmetry breaking is increased. This decreases $SU(3)$
type contributions to (\ref{mnvm}).}.

The addition of vector mesons also plays an important role in the
discussion of the ``sizes" of the nucleons: the nucleon radii.
As can be observed from table \ref{ta_emps} the Skyrme model of
pseudoscalars only seriously
underestimates the empirical values for the baryon radii.
The presence of the $\omega$ meson provides an increase of
the isoscalar radius \cite{Me87}
\be
\langle r^2\rangle _{I=0} \approx
\langle r^2\rangle _B +\frac{6}{m_\rho^2} \ ,
\label{changeR0}
\ee
where $\langle r^2\rangle _B$ is the radius associated with baryon
number current (\ref{topcur}). The additional piece in eq (\ref{changeR0}) 
is a consequence of (approximate) vector meson dominance in this model
\cite{Sch86}, which 
indeed  is observed when including the vector mesons in a chirally 
invariant manner. As can be seen from table \ref{ta_emps} this
increase of about $0.35{\rm fm}^2$ will significantly improve the
predictions for the radii.

A similar interesting improvement due to vector mesons is obtained
in the context of meson--baryon scattering. In these investigations
one introduces small fluctuations off the classical soliton. Eventually
these fluctuations are quantized to represent in-- and out--going
meson fields, thereby determining the scattering matrix \cite{Wa84}.
It turns out that in the pseudoscalar Skyrme model the phase--shifts
extracted from this scattering matrix rise almost linearly with 
the momentum of the in--going pion. This undesired feature is mostly
due to the contact interaction between pions contained in the 
Skyrme model Lagrangian ({\it cf.} section 2). When introducing
vector mesons this contact interaction is essentially replaced by the
exchange
of such a vector meson,
\be
\frac{-1}{m_\rho^2}\, \longrightarrow\, 
\frac{1}{q^2-m_\rho^2}\, .
\label{vmscat}
\ee
As this interaction decreases for large momentum transfers, $q^2$ the
resulting phase--shifts assume a constant value for large energies
rather than rising linearly \cite{Schw87}. Clearly this effect is 
similar to the one observed when going from the Fermi to the standard 
model of electro--weak interactions.

These examples show that while the inclusion of vector meson degrees 
of freedom involves quite a few technical details it clearly provides 
a more realistic picture of the nucleon as a chiral soliton.

\bigskip
\stepcounter{chapter}
\leftline{\Large \bf 6. Summary and discussion}
\medskip

Aside from the mass spectra and current matrix elements of the 
low--lying $\frac{1}{2}^+$ and $\frac{3}{2}^+$ baryons treated here 
the soliton approach has been extensively employed to study meson 
nucleon scattering \cite{Wa84,Schw87}, baryons containing a 
heavy quark \cite{Ca85}, nucleon--nucleon scattering \cite{Ja85},
few nucleon systems \cite{We86} and nuclear matter \cite{Ja88a}.

In the present survey, we started out with a historical introduction
(section 1) and a concise technical summary of the original two flavor
Skyrme model (section 2).  In these sections the physical interpretation
and justification of the model were emphasized; it is hoped that they will
be useful to beginners in this area of research (see also \cite{Ho86}
and \cite{Za86}).

We next attempted to develop the generalization of the original Skyrme
model which is suggested by the large $N_C$ approximation to QCD.
In this approach the Skyrme Lagrangian is to be replaced by a more
general effective Lagrangian containing mesons of all spins. Perhaps some
day an analytic expression in this framework will be found. At present it
seems necessary to obtain an approximation based on including the lowest
energy resonances and constraining the model by the symmetries of the
underlying QCD. The concept of chiral symmetry which plays a crucial role
in this extension was explored in section 3. Furthermore the original
Skyrme model of two light flavors was extended to three flavors, as
it is now well established that the nucleons belong to a flavor SU(3)
multiplet.

It is worthwhile to stress that once the effective Lagrangian has been
determined from the meson sector, the soliton approach provides in 
principle a {\it zero parameter} description of baryon properties 
(In our case we introduced just one parameter which had to be fit 
from the baryon sector.).

In section 4 we studied the technical tools needed to treat the flavor
SU(3) symmetry and its breaking at the (collective) baryon level. These
were applied to the calculation of various interesting baryon 
matrix elements. Finally section 5 sketched the treatment of baryons 
based on an effective Lagrangian which also included the vector mesons. 
An application to the so--called {\it proton spin puzzle} demonstrated 
that the soliton approach seems to give a neat description of, otherwise 
hard to explain, experimental results on the quark spin structure of 
the nucleon. The improvements one encounters on including the vector 
mesons are in accord with the intuitive notion that the addition of 
higher mass resonances in the meson sector leads to a progressively 
more detailed understanding of the short distance structure of the 
nucleon--as--soliton.

We are happy to acknowledge the stimulating interactions we have had
with many collaborators and colleagues while doing research related
to the topics reviewed here.

\end{document}